\documentclass[aip, pof,graphicx]{revtex4-2}

\usepackage{graphicx}
\usepackage{dcolumn}
\usepackage{bm}
\usepackage[utf8]{inputenc}
\usepackage[T1]{fontenc}
\usepackage{color}
\usepackage{longtable}
\usepackage{amssymb}
\usepackage{epstopdf}
\newcommand{\corr}[1]{{\color{black} #1}}

\draft 

\begin{document}


\title[Forces and grain-grain contacts in bidisperse beds sheared by viscous fluids]{Forces and grain-grain contacts in bidisperse beds sheared by viscous fluids\\
	This article appeared in Phys. Fluids 36, 113342 (2024) and may be found at https://doi.org/10.1063/5.0238582.}



\author{Jaime O. Gonzalez}
\altaffiliation[Also at ]{Faculdade de Engenharia Mec\^anica, Universidade Estadual de Campinas (UNICAMP),\\
	Rua Mendeleyev, 200, Campinas, SP, Brazil}
\affiliation{Departamento de Petróleos, Escuela Politécnica Nacional,\\
	Av. Ladrón de Guevara E11-253, Quito, Ecuador
}%

\author{Erick M. Franklin*}%
 \email{erick.franklin@unicamp.br}
 \thanks{*Corresponding author}
\affiliation{Faculdade de Engenharia Mec\^anica, Universidade Estadual de Campinas (UNICAMP),\\
Rua Mendeleyev, 200, Campinas, SP, Brazil
}%


\date{\today}

\begin{abstract}
In a recent paper (Gonzalez et al., 2023), we investigated the motion of grains within a granular bed sheared by a viscous fluid, and showed how segregation and hardening occur in the fluid- (bedload) and solid-like (creep) regions. In this paper, we inquire further into the mechanisms leading to grain segregation in a bidisperse bed, and how the forces are distributed. For that, we carried out numerical simulations at the grain scale by using CFD-DEM (computational fluid dynamics-discrete element method), with which we were able to track the positions, velocities, forces, and solid contacts underwent by each grain. We show that during the upward motion of large grains the direct action of fluid forces is significant in the middle and upper parts of the bedload layer, while only contact forces are significant in the creep layer and lower part of the bedload layer. We also show that in all cases the particles experience a moment about a -45$^\circ$ contact point (with respect to the horizontal plane) when  migrating upward, whether entrained by other contacts or directly by the fluid. In addition, we show the variations in the average solid-solid contacts, and how forces caused either by solid-solid contacts or directly by the fluid are distributed within the bed. Our results provide the relationship between force propagation and reorganization of grains in sheared beds, explaining mechanisms found, for example, in river beds and landslides.

\end{abstract}


\maketitle 

\section{INTRODUCTION}
\label{sec:intro}

When a granular bed is sheared by a liquid flowing under moderate shear stresses (shear forces comparable to the grains' weight), grains are transported as bedload: a moving layer in which grains are entrained by rolling and sliding over each other \cite{Bagnold_1, Raudkivi_1}. The bedload layer moves over a creep layer, where small and slow rearrangements of grains deform this part of the bed \cite{Houssais_1, Gonzalez}. Because the creep layer is deformed over long times, while the bedload layer presents a durable and high rate of deformation, they are described as solid- and fluid-like layers, respectively \cite{Houssais_1, Gonzalez}.

Bedload in liquids is known to take place after a given threshold of the shear stress is surpassed, since below that threshold the fluid drag is not strong enough for causing grains to roll over one another (promoting motion). For conditions below that threshold, the bed has been described, until recently, as static. In addition, because shear stresses of the liquid decrease to the threshold value within the bedload layer (evaluating the shear stress profile from top to bottom), the layer beneath bedload has also been described as static \cite{Raudkivi_1, Yalin_1}. However, Houssais et al. \cite{Houssais_1} and Allen and Kudrolli \cite{Allen_2} showed that that portion of the bed creeps, and that this creeping layer can exist even when shear stresses are below the threshold for bedload, i.e., without relevant grain transport (there are only rearrangements of grains). Investigating further the motion of sheared beds, Houssais et al. \cite{Houssais_1} showed that there is a continuous transition between bedload and creep, and that the transition occurs at a height characterized by the ratio between the microscopic (related to the rearrangements of grains) and macroscopic (related to the macroscopic rate of deformation) timescales. In the case of viscous liquids this ratio corresponds to the viscous number \cite{Jop} $I_v$, and Houssais et al. \cite{Houssais_1} found that transition occurs at the height for which $I_v$ $=$ 10$^{-7}$.

The threshold for bedload is not constant, but varies over time \cite{Charru_1}. Under constant flow velocities, the threshold stress increases with time, implying the decrease in grains' mobility. Moreover, in cases of small applied stresses it can stop completely bedload from a certain time on. The decrease in granular mobility over time is known as bed hardening, and it is caused by (i) the reorganization of grains within the bed (happening in both mono and polydisperse beds); and, (ii) the armoring due to particle segregation in polydisperse beds, i.e., an increase in the concentration of larger particles on the bed surface, sheltering from the fluid flow the smaller (and more mobile) particles. 

For the reorganization of grains in a monodisperse bed, we can identify isotropic and anisotropic contributions. The isotropic contribution is that caused by purely geometric rearrangements of grains, in which grains percolate within the bed toward vacancies, leading, thus, to an increase in bed compactness. This idea has been proposed, for example, by Charru et al. \cite{Charru_1} and Masteller and Finnegan \cite{Masteller_1}, who carried out flume experiments in which they measured the decay in the mobility of a bedload layer. This decay was also measured in the field by Masteller et al. \cite{Masteller_2}, who identified hardening in the bed of the Erlenbach river by analyzing a 19-year-series dataset of fluid stress and sediment transport. However, the explanation of bed hardening caused only by percolation does not suffice: there is also an anisotropic contribution due to the organization of grains in preferential directions, forming chains that percolate forces by solid-solid contacts. The formation of those force chains, and transmission of forces through a contact force network, was initially studied in the context of static or quasi-static granular matter \cite{Cates, Majmudar, Bi}. In particular, Cates et al. \cite{Cates} showed that fragile states where force chains are aligned in preferential directions can appear, so that a granular matter in those states jams and supports loading in such directions, but not in others, while Bi et al. \cite{Bi} showed that granular matter is prone to fragile states and shear jamming (when external shear stresses are applied) at particle fractions lower than those necessary for jamming by an isotropic compression. Later, C\'u\~nez et al. \cite{Cunez2} investigated experimentally the time evolution of the structure of a granular bed sheared by a viscous fluid. In their experiments, the bed was submitted to different shear cycles, with and without variations in direction, in order to inquire into the formation and breakage of contact chains. They showed that, when sheared by a viscous flow in a constant direction during a certain period, an anisotropic structure appears, but this structure is broken by reversing the flow direction, while the bed compaction, which corresponds to the isotropic hardening, is not affected. This implies a decay in sediment transport if the fluid flow is sustained in one direction, and the recovery of a portion, corresponding to the anisotropic structure, if the flow is reversed. The portion due to compaction is only recovered if the shear is strong enough for dilating the bed. 

Concerning polydisperse beds, the migration of larger particles toward the bed surface leads to size armoring: smaller particles become sheltered from regions where the flow velocities are higher \cite{Frey}. For this reason, some of previous studies investigated grain segregation in bidisperse bedload and debris flow \cite{Zhou2, Cui, Rousseau, Frey2, Ferdowsi, Gonzalez}. In particular, Rousseau et al. \cite{Rousseau} investigated the upward motion of a single particle inserted in a bedload layer consisting of smaller grains, and showed that its upward motion has two distinct phases: an intermittent and slow motion from deeper regions within the bedload, followed by a fast motion to the bed surface (later, Gonzalez et al. \cite{Gonzalez} conjectured that the first phase corresponds to the limit between the creep and bedload layers). Ferdowsi et al. \cite{Ferdowsi} investigated segregation and hardening in a bidisperse bed sheared by a steady viscous flow by carrying out experiments in which the ratio between the solid and fluid densities $S$ was 1.13 (close to unity), while in nature $S$ $\approx$ 2.65 (water and sand). In addition, they performed numerical simulations for the grains only (without fluid), in which the particles were sheared by an imposed boundary condition. They found that in deeper regions creep leads to a slow segregation in a diffusion-like manner, while closer to the surface bedload leads to a fast, shear-dependent segregation, in agreement with the results of Rousseau et al. \cite{Rousseau}. \corr{Both Zhou et al. \cite{Zhou2} and Cui et al. \cite{Cui} investigated the effect of an interstitial fluid in particle segregation, the former for debris flows and the latter for shear flows, and found that segregation decreases with increasing the fluid viscosity. They proposed that the fluid diminishes particle-particle contacts and dampens particle fluctuations (an effect that increases with viscosity), weakening, thus, particle segregation.} In a recent paper \cite{Gonzalez}, we investigated experimentally bed hardening and particle segregation in a bidisperse bed sheared by a viscous liquid, for the case $S$ = 2.7 (close to values found in nature). Each experimental run was carried out for 140 hours and we used refractive index matching, allowing measurements of both bedload and creep in a plane far from the channel walls (a plane inside the bed). Among other findings, we showed that there exist diffusive, advective and constrained regions, that segregation occurs within the bedload layer and in the 5\% topmost region of the creep layer (at least for the total duration of our experiments), that most of segregation occurs during the very first stages of the flow, that bed hardening becomes stronger while bedload and creep weaken along time, and we proposed characteristic times for segregation and hardening. In our experiments, the creep-bedload transition occurred at  $I_v$ $\approx$ 2 $\times$ 10$^{-8}$, instead of 10$^{-7}$.

\corr{For the moment, to the best of our knowledge, there is not a real understanding of the forces driving the larger particles upward. For instance, mechanisms of kinetic sieving, for which an interstitial fluid would hinder segregation, have been proposed, but no measurements of the solid-solid contact forces and of the fluid forces acting on each grain have been reported. Measurements of the forces caused by the fluid would help understanding its real role (of promoting or hindering segregation), while those of contact forces would explain if there is kinetic sieving or other mechanism pushing large grains upward. In addition, the knowledge of the magnitude and direction of forces would show if vertical forces act directly on the segregating particle, or if horizontal components make then roll over specific contact points (moving, thus, upward).}

The previous studies have advanced our knowledge on segregation and hardening, but important questions on the mechanics of hardening and segregation remain, however, to \corr{be} answered. For example, what are the magnitudes of forces acting on grains being segregated? What are the relative contributions of forces due directly to the fluid and to solid-solid contacts? How are solid-solid contacts distributed around segregating particles, and how they evolve as those particles move upward? And how the network of contact forces varies along time? In this paper, we aim to answer these questions by carrying out numerical simulations at the grain scale of a bidisperse bed sheared by a viscous fluid. We made use of CFD-DEM (computational fluid dynamics-discrete element method), and, because we computed the motion of each grain at all times, we tracked the positions, velocities, forces, and solid contacts underwent by each one of them. For the larger grains moving upward, we show that the direct action of fluid forces is significant in the middle and upper parts of the bedload layer, while only contact forces are significant in the creep layer and lower part of the bedload layer. We also show that those grains experience a moment about a -45$^\circ$ contact point (with respect to the horizontal plane) whether entrained by other contacts or directly by the fluid. In addition, we present the variations in the average solid-solid contacts, and how forces caused either directly by the fluid or solid-solid contacts are distributed within the bed.

In the following, Sec. \ref{sec:setup} presents the fundamental equations and numerical setup, Sec. \ref{sec:results} shows the results and discussion, and Sec. \ref{sec:conclusions} concludes the paper.

\section{Basic equations and numerical setup}
\label{sec:setup}

\subsection{Model equations}

We carried out CFD-DEM computations, in which the fluid flow was in laminar regime (as shown next, the Reynolds numbers of the channel flow were smaller than unity), by using the open-source CFD and DEM codes OpenFOAM and LIGGGHTS \cite{Kloss, Berger}, respectively, coupled by the open-source code \mbox{CFDEM} (www.cfdem.com) \cite{Goniva}. The DEM part solves the linear (Eq. \ref{Fp}) and angular (Eq. \ref{Tp}) momentum equations for each solid particle in a Lagrangian framework,

\begin{equation}
m_{p}\frac{d\vec{u}_{p}}{dt}= \vec{F}_{p}\,\, ,
\label{Fp}
\end{equation}

\begin{equation}
I_{p}\frac{d\vec{\omega}_{p}}{dt}=\vec{T}_{c}\,\, ,
\label{Tp}
\end{equation}

\noindent where, for each grain, $m_{p}$ is the mass, $\vec{u}_{p}$ is the velocity, $I_{p}$ is the moment of inertia, $\vec{\omega}_{p}$ is the angular velocity, $\vec{T}_{c}$ is the resultant of contact torques between solids, and $\vec{F}_{p}$ is the resultant force,

\begin{equation}
\vec{F}_{p}= \vec{F}_{fp} + \vec{F}_{c} + m_{p}\vec{g}\,\, .
\label{Fp2}
\end{equation}

\noindent In Eq. \ref{Fp2}, $\vec{g}$ is the acceleration of gravity, $\vec{F}_{c}$ is the resultant of contact forces between solids, and $\vec{F}_{fp}$ is the resultant of fluid forces acting on each grain, 

\begin{equation}
	\vec{F}_{fp} = \vec{F}_{d} + \vec{F}_{p} + \vec{F}_{\tau} + \vec{F}_{vm} \,\, .
	\label{Ffp_sim}
\end{equation}

\noindent In Eq. \ref{Ffp_sim}, we considered the contributions due to fluid drag, $\vec{F}_{d}$, pressure gradient, $\vec{F}_{p}$, gradient of the deviatoric stress tensor, $\vec{F}_{\tau}$, and virtual mass, $\vec{F}_{vm}$, and neglected the Basset, Saffman and Magnus forces because they are usually of lesser importance for CFD-DEM simulations \cite{Zhou}. In Eq. \ref{Tp}, we neglect the torques caused directly by the fluid since they are considered much smaller than those due to contacts \cite{Tsuji, Tsuji2, Liu}. The contact forces $\vec{F}_{c}$ and torques $\vec{T}_{c}$ are computed by Eqs. \ref{Fc} and \ref{Tc}, respectively,

\begin{equation}
	\vec{F}_{c} = \sum_{i\neq j}^{N_c} \left(\vec{F}_{c,ij} \right) + \sum_{i}^{N_w} \left( \vec{F}_{c,iw} \right) \,\, ,
	\label{Fc}
\end{equation}

\begin{equation}
	\vec{T}_{c} = \sum_{i\neq j}^{N_c} \vec{T}_{c,ij} + \sum_{i}^{N_w} \vec{T}_{c,iw} \,\, ,
	\label{Tc}
\end{equation}

\noindent where $\vec{F}_{c,ij}$ and $\vec{F}_{c,iw}$ are the contact forces between particles $i$ and $j$ and between particle $i$ and the wall, respectively, $\vec{T}_{c,ij}$ is the torque due to the tangential component of the contact force between particles $i$ and $j$, and $\vec{T}_{c,iw}$ is the torque due to the tangential component of the contact force between particle $i$ and the wall. $N_c$ $-$ 1 is the number of particles in contact with particle $i$, and $N_w$ is the number of particles in contact with the wall. In Eqs. \ref{Fc} and \ref{Tc}, we apply the Hertzian model, which is described briefly in the supplementary material and in more detail in Ref. \cite{Lima2}.

The CFD part computes the mass and momentum equations for the fluid in an Eulerian framework. In our simulations, we use an unresolved approach where the equations are phase-averaged in a volume basis while assuring mass conservation, resulting in Eq. \ref{mass_fluid} for the mass and Eq. \ref{momentum_fluid} for momentum,

\begin{equation}
	\frac{\partial \left( \alpha_{f} \rho_{f} \right)}{\partial t} + \nabla \cdot \left ( \alpha_{f} \rho_{f} \vec{u}_{f} \right ) = 0 \,\,,
	\label{mass_fluid}
\end{equation}

\begin{equation}
	\frac{\partial \left ( \alpha_{f} \rho_{f} \vec{u}_{f} \right ) }{\partial t} + \nabla \cdot \left ( \alpha_{f} \rho_{f} \vec{u}_{f} \vec{u}_{f} \right ) = -\alpha_{f} \nabla P - \vec{f}_{exch} + \alpha_{f} \nabla \cdot  \vec{\vec{\tau}}_{f}  + \alpha_{f} \rho_{f} \vec{g} \,\,,
	\label{momentum_fluid}
\end{equation}

\noindent where $\alpha_{f}$ is the volume fraction of the fluid, $\vec{u}_{f}$ is the fluid velocity, $\rho_{f}$ is the fluid density, $P$ is the fluid pressure, $\vec{\vec{\tau}}$ is the deviatoric stress tensor of the fluid, and $\vec{f}_{exch}$ is an exchange term between the grains and the fluid. The latter is different from $\vec{F}_{fp}$ divided by the volume because, during the averaging process for obtaining Eqs. \ref{mass_fluid} and \ref{momentum_fluid}, the forces due to the pressure gradient and deviatoric stress tensor were split from the remaining fluid-particle forces. Since we neglect Basset, Saffman and Magnus forces,

\begin{equation}
	\vec{f}_{exch} = \frac{1}{\Delta V}\sum_{i}^{n_{p}} \left( \vec{F}_{d} +  \vec{F}_{vm} \right) \,\, ,
	\label{forces_exchange}
\end{equation}

\noindent where $n_p$ is the number of particles in the considered cell of volume $\Delta V$. More details on the computation of $\vec{f}_{exch}$ are available in Lima et al. \cite{Lima2}.

\subsection{Numerical setup}

\begin{figure}[h!]
	\begin{center}
		\includegraphics[width=.6\linewidth]{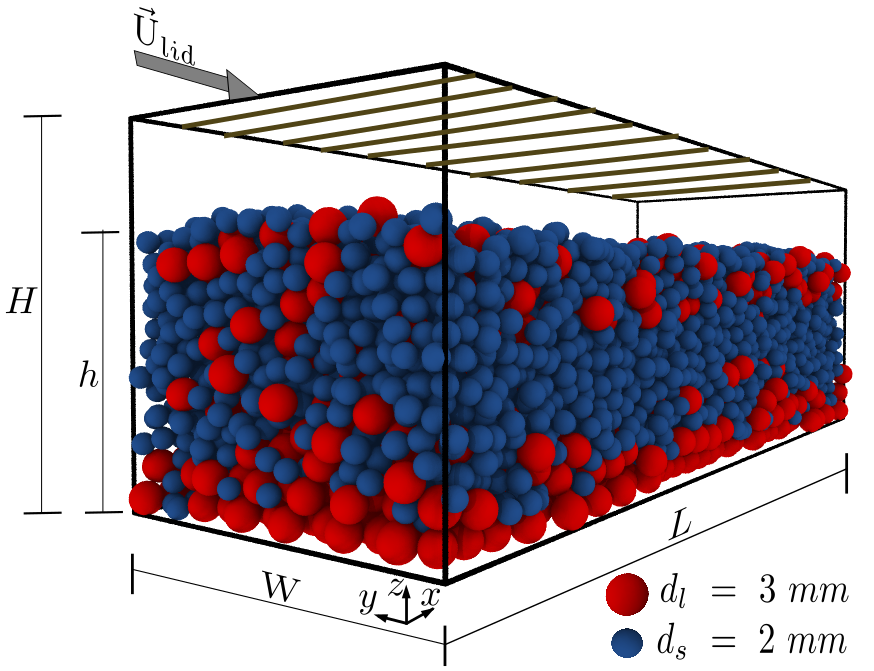}\\
	\end{center}
	\caption{Layout of the numerical setup, where $L$ = 100 mm, $W$ = 40 mm, and $H$ = 30 mm.}
	\label{fig:setup}
\end{figure}

The numerical setup consisted basically of a rectangular box filled with bidisperse grains and a viscous fluid, the latter sheared by an imposed constant velocity at the top surface (top boundary), and periodic conditions are imposed at the inlet and outlet (i.e., in the longitudinal direction). The used box can be seen as a small section of the  annular (circular) flume with rotating lid of our previous experiments, reported in Gonzalez et al. \cite{Gonzalez}. All other parameters are similar to those used in the experiments \cite{Gonzalez}: the periodic channel had $W$ = 40 mm width and $H$ = 30 mm height, being completely filled with grains and a viscous liquid. In order to emulate the experiments, the fluid velocity at the top boundary varied within 53 mm/s $\leq$ $U_{lid}$ $\leq$ 106 mm/s.

Prior to simulations, two different populations of solid spheres were distributed in the domain with a layer of larger particles below another of smaller particles (within each layer, the corresponding species was randomly distributed), and let to fall freely until settling at the bottom. After an initial re-suspension (described next), they formed a bidisperse granular bed of height 24 mm $\leq$ $h$ $\leq$ 25 mm. For the grain populations, we used spheres with density $\rho_p$ = 2500 kg/m$^3$ and diameters $d_l$ = 3 mm $\pm$ 0.2 mm and $d_s$ = 2 mm $\pm$ 0.2 mm, which we call species $l$ and $s$ (standing for large and small), respectively. Both species were produced with sizes following a Gaussian distribution with standard deviation equal to 0.1 mm. As in Gonzalez et al. \cite{Gonzalez}, the ratio of the total volume occupied by the small spheres ($V_s$) to that of large ones ($V_l$) was $V_s/V_l$ = 1.5, and we consider the mean diameter as $d$ $=$ $(0.4d_l + 0.6d_s) /2$ $=$ 2.4 mm. \corr{The 3:2 ratio of particle diameters was chosen to match the value used in our previous experiments \cite{Gonzalez}, which we know to promote considerable segregation while avoiding excessive large ratios.} The physical properties of grains correspond to glass particles, where for the coefficient of restitution $e$ we considered the fact that simulations are unresolved and the interstitial fluid is a viscous liquid\corr{, i.e., $e$ embeds the effects of the lubrication force. This procedure is corroborated by experiments with bouncing particles, such as those of Gondret et al. \cite{Gondret}}. The used values are listed in Tab. \ref{properties_grains}.

\begin{table}[!h]
	\centering
	\caption{Physical properties of solid spheres used in the simulation.}
	\begin{tabular}{c|c}
		\hline
		Parameter & Value \\ \hline
		Sliding Friction Coeff. $\mu$   & 0.6  \\
		Rolling Friction Coeff. $\mu_r$  & 0   \\
		Restitution Coef. $e$   & 0.051  \\
		Poisson Ratio $\sigma$   & 0.245   \\
		Young's Modulus $E$ (GPa) & 0.46    \\
		Density $\rho_p$ (kg/m$^3$)  & 2500  \\ \hline		
	\end{tabular}
	\label{properties_grains}
\end{table}

For the fluid, we used a liquid with dynamic viscosity $\eta$ = 651 cP and density $\rho_f$ = 931 kg/m$^3$. Because the granular bed settled on the bottom, there remained a liquid film of height 5.3 mm $\leq$ $h_f$ $\leq$ 5.8 mm above the bed, which was sheared by imposing the velocity $U_{lid}$ at the top boundary. With that, the shear rates, $\dot{\gamma}_f$ = $U_{lid,r}/h_f$, varied within 5.7 and 11.3 s$^{-1}$, the Reynolds number based on the fluid height, Re = $\rho_f U_{lid} h_f \eta^{-1}$, within 0.42 and 0.70, the Reynolds number based on the mean grain diameter, Re$_{p}$ = $\rho_f \dot{\gamma}_f d^2 \eta^{-1}$, within 0.043 and 0.093, and the Shields number, $\theta$ = $\eta \dot{\gamma}_f ((\rho_p - \rho_f )g d)^{-1}$, within 0.15 and 0.32, where $g$ = $|\vec{g}|$ and $U_{lid,r}$ = $U_{lid} - \left< u_{p,surf} \right>$. In its turn, $\left< u_{p,surf} \right>$ is the space-time average of grains on the bed surface. The ratio between the grain and fluid densities is $S$ = $\rho_p / \rho_f$ = 2.7, very close to values found in nature. As in Gonzalez et al. \cite{Gonzalez}, we imposed a high velocity ($U_{lid}$ = 534 mm/s) at the top boundary prior to simulations, reaching $\theta$ = 1.5 during 21 s, which assured that all grains with the exception of the bottom-most layers were suspended, with some partial segregation being unavoidable. This was followed by a period of rest in which the grains settled, and by the end of that time we reached the initial condition used in all tests. Only after that, the desired value of $U_{lid}$ was imposed and each simulation began.

The numerical domain is a channel of size $L_x$ = 100 mm, $L_y$ = $W$ = 40 mm and $L_z$ = $H$ = 30 mm, where $x$, $y$ and $z$ are the longitudinal, spanwise, and vertical directions, respectively, and the mesh used for the fluid consisted of 10080 hexaedral cells, with a maximum cell volume of approximately 1.65 $\times$ 10$^{-8}$ m$^3$ (a layout of the mesh is shown in the supplementary material). \corr{This corresponds to a ratio between the cell size and particle diameter of 0.8 and 1.3 for the large and small particles, respectively. A table with a GCI (grid convergence index) test \cite{roache1998verification} for a pure oil flow is available in the supplementary material, from which we notice that the employed mesh has enough refinement}. For the fluid, the boundary conditions are periodic flow in the longitudinal direction, impermeability and no-slip conditions at the bottom and lateral walls, and imposed velocity at the top boundary. For the grains, the boundary conditions are solid wall at the walls and periodic conditions in the longitudinal direction. The time step used for the DEM was 10$^{-6}$ s, which was less than 20 \% of the Rayleigh time \cite{Derakhshani} in all simulated cases, and that of CFD was 10$^{-4}$ s, which respected the CFL (Courant-Friedrichs-Lewy) criterion \cite{Courant}. The total duration of simulated cases was 20 min (real time). Figure \ref{fig:setup} shows a layout of the numerical setup, and more details are available in an open repository \cite{Supplemental, Supplemental2}.

\section{RESULTS AND DISCUSSION}
\label{sec:results}

\begin{figure}[h!]
	\begin{center}
		\includegraphics[width=.85\linewidth]{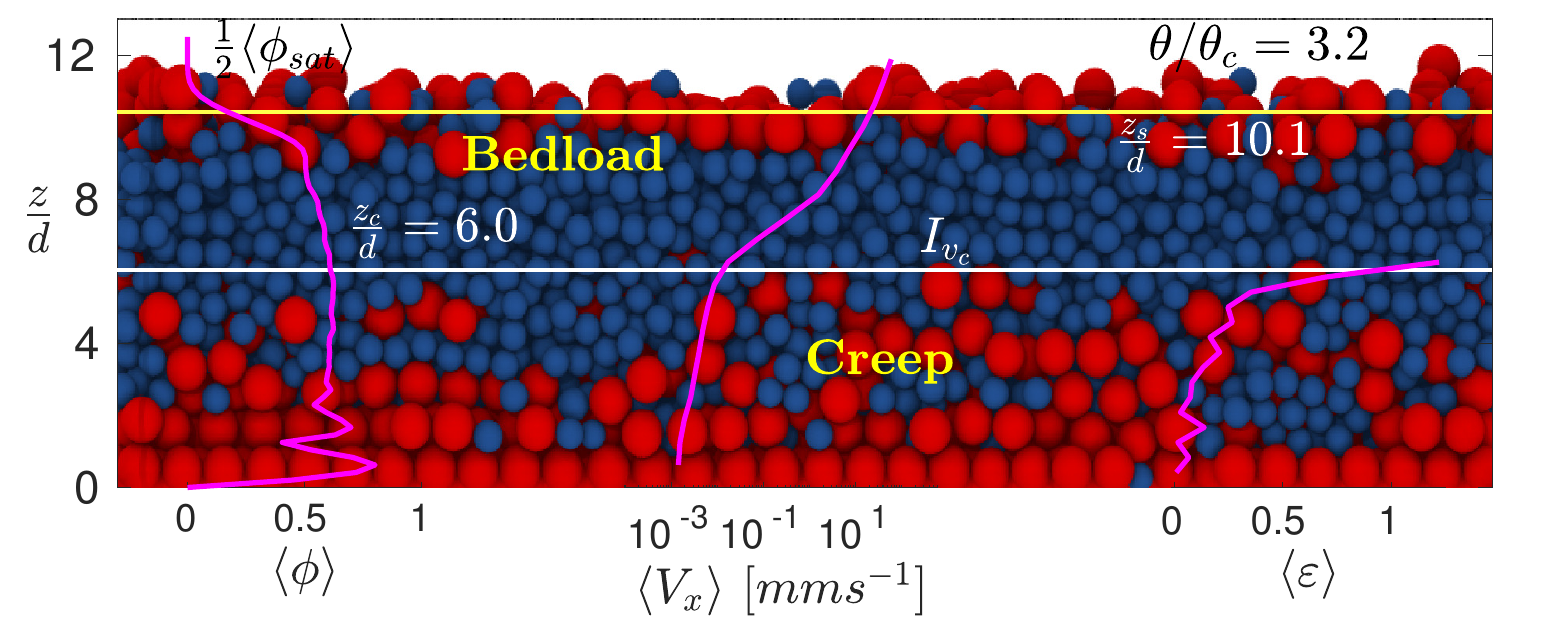}\\
	\end{center}
	\caption{Space-time averages of the packing fraction $\left< \phi \right>$, longitudinal velocity $\left< V_x \right>$, and strain $\left< \varepsilon \right>$, for $\theta / \theta_c$ = 3.2, superposed with the final positions of grains. In the image, the flow is from left to right.}
	\label{fig:bed_general}
\end{figure}

The bed dynamics was similar to that observed in experiments \cite{Gonzalez}, the fluid flow entraining the solid spheres (grains) into motion with two distinct layers: a bedload layer above a creep layer. The bedload layer has a fluid-like behavior in which the grains move by rolling and sliding over each other, while the creep layer presents a solid-like behavior in which the grains have much smaller velocities, resulting in a slow deformation of this layer. Because in our simulations we have access to instantaneous positions of all grains, we can compute not only mean quantities such as space-time averages within the bed of the packing fraction $\left< \phi \right>$, longitudinal velocity $\left< V_x \right>$, and strain $\left< \varepsilon \right>$, but also the trajectories of grains and the forces (solid-solid, fluid-solid, and resultant) acting on each grain, which we present next. Figure \ref{fig:bed_general} shows the space-time averages of the packing fraction $\left< \phi \right>$, longitudinal component of velocity $\left< V_x \right>$, and strain $\left< \varepsilon \right>$, superposed with the final positions of grains, for $\theta / \theta_c$ = 3.2. The Shields number is normalized by the critical value $\theta_c$ = 0.1, which corresponds to the threshold of bedload \cite{Houssais_1, Houssais_2, Ferdowsi}. Space averages were computed only in the longitudinal ($x$) and spanwise ($y$) directions, and the agreement with the experiments of Gonzalez et al. \cite{Gonzalez} is excellent.

\subsection{Bed profiles}

\begin{figure}[h!]
	\begin{center}
		\includegraphics[width=.95\linewidth]{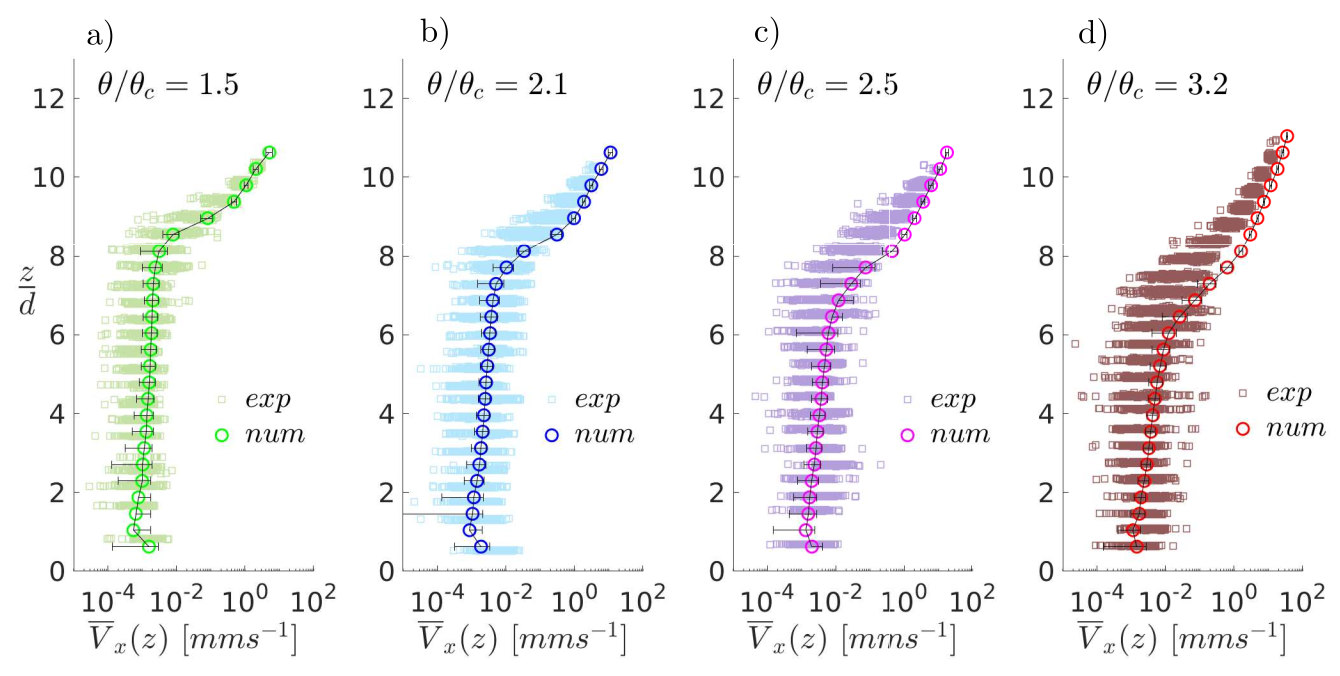}\\
	\end{center}
	\caption{(a)-(d): Space averages of the longitudinal velocity within the bed $\overline{V}_x(z)$, for all detected grains and different shear stresses $\theta / \theta_c$, for the present simulations and the experiments of Gonzalez et al. \cite{Gonzalez}. Symbols are shown in the key of each panel, where for the simulations we used the mean value and error bars, while for the experiments we plotted all values using symbols.}
	\label{fig:bed_velocities}
\end{figure}

Figure \ref{fig:bed_velocities} shows the vertical profiles of the longitudinal component of the space-averaged velocity $\overline{V}_x$, for both the present results and the experiments of Gonzalez et al. \cite{Gonzalez}. In Fig. \ref{fig:bed_velocities}, the experimental values for the first 20 min are shown as squares, and the numerical values are shown as circles (time-mean values) and error bars. For the numerical results, the values of $\overline{V}_x(z)$ were obtained by averaging the longitudinal component of $\vec{u}_{p}$ along the $x$ and $y$ axes (in the case of experiments, they were considered in the plane of measurement, i.e., only in the $x$ axis). We notice that velocities are higher on the bed surface and decrease relatively fast within the bedload layer (6-8 $\leq$ $z/d$ $<$ 10-12), having a smoother decrease and reaching much lower values (5 orders of magnitude smaller than on the surface) in the creep layer ($z/d$ $<$ 6-8). By considering the dispersion of the experimental data, we observe a good agreement between the velocity profiles obtained from numerical computations and from  experiments. In particular, we observe the kink in the profiles \cite{Houssais_1, Houssais_2}, separating the bedload from the creep region.

We also computed the vertical profiles of the viscous number $I_v$, confinement pressure $P_p$, and effective viscosity $\eta_{eff}$. The viscous number $I_v$ represents the ratio between the microscopic (rearrangement of grains) and macroscopic (rate of deformation of the bed) timescales \cite{GDR_midi}, and is given by Eq. \ref{eq:Iv},

\begin{equation}
I_v = \frac{\eta \dot{\gamma}}{P_p} \,\,,
\label{eq:Iv}
\end{equation}

\noindent where $\dot{\gamma}$ is the shear rate of the granular material. We computed the confinement $P_p$ as in Houssais et al. \cite{Houssais_2},

\begin{equation}
	P_p = \left( \rho_s - \rho \right) g \left[ \frac{\forall_s}{A_{cont}} +  \int_{z}^{\infty} \left< \phi \right> dz \right] \,\,,
	\label{eq:Pp}
\end{equation}

\noindent where $\forall_s$ is the volume of one grain (for which we consider the mean diameter $d$) and $A_{cont}$ is the characteristic surface of contact between the topmost grain and the bed surface. According to Houssais et al. \cite{Houssais_2}, $\forall_s / A_{cont}$ = 0.1$d$, which we used in our computations. The effective viscosity $\eta_{eff}$ is computed by Eq. \ref{eq:eta_eff},

\begin{equation}
	\eta_{eff} = \frac{\tau}{\dot{\gamma}} \,\,,
	\label{eq:eta_eff}
\end{equation}

\noindent where $\tau$ is the norm of the shear stress. \corr{The effective viscosity $\eta_{eff}$ is important for continuum models, where $\eta_{eff}$ is modeled as a function of the viscous number $I_v$ (or the inertial number, depending on the problem).}

\begin{figure}[h!]
	\begin{center}
		\includegraphics[width=.85\linewidth]{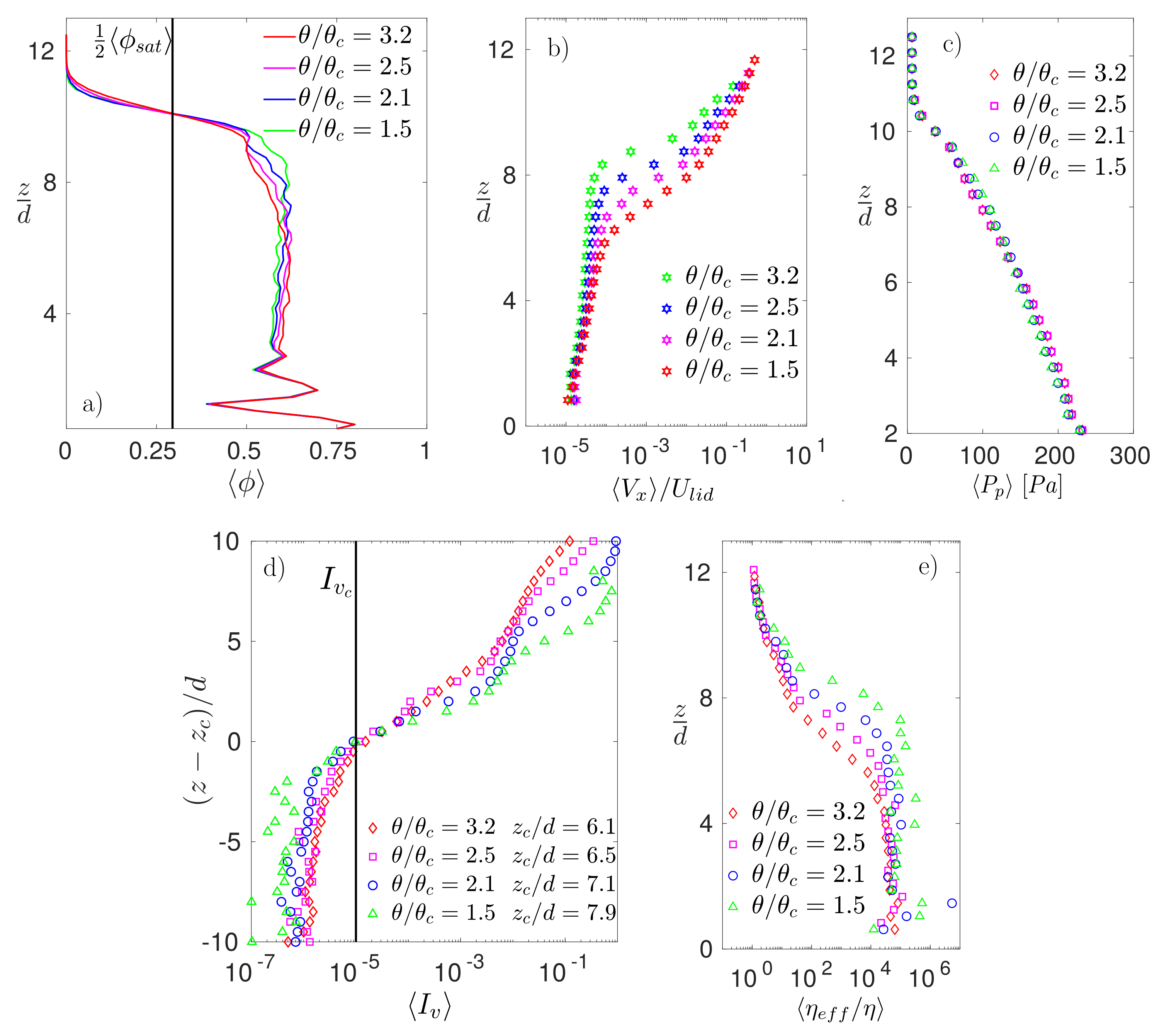}\\
	\end{center}
	\caption{Vertical profiles of space-time averaged (a) packing fraction, $\left< \phi \right>$, (b) grain velocity in the longitudinal direction, $\left< V_x \right>$, normalized by that of the lid, $U_{lid}$, (c) confinement pressure, $\left< P_p \right>$, (d) viscous number, $\left< I_v \right>$, and (e) effective viscosity $\left< \eta_{eff} \right>$ normalized by that of the fluid, $\eta$, for different shear stresses $\theta / \theta_c$. Space averages are computed only in the longitudinal and spanwise directions, and in panel (d), the vertical coordinate has its origin in the creep-bedload transition.}
	\label{fig:profiles}
\end{figure}

Figures \ref{fig:profiles}a--\ref{fig:profiles}e show the vertical profiles of the space-time averaged packing fraction $\left< \phi \right>$, longitudinal component of the grain velocity $\left< V_x \right>$, confinement pressure $\left< P_p \right>$, viscous number $\left< I_v \right>$, and effective viscosity $\left< \eta_{eff} \right>$, for different shear stresses $\theta / \theta_c$. If we consider the first 20 min of the Gonzalez et al. \cite{Gonzalez} data, the agreement is excellent. In particular, we observe in Fig. \ref{fig:profiles}d that the kink in velocity corresponds to $I_v$ $\sim$ $\times$ 10$^{-5}$, higher than the value found by Gonzalez et al. \cite{Gonzalez} in the end of experiments. This happens because during the first 20 min the bed has not enough time to harden considerably (the experiments of Gonzalez et al. \cite{Gonzalez} took nearly 6 days, a duration prohibitive in the present simulations). The corresponding graphics plotted with the first 20 min of the Gonzalez et al. \cite{Gonzalez} data are available in the supplementary material (they can be compared directly with the numerical results). Finally, as in Refs. \cite{Houssais_1, Cunez2, Gonzalez}, the vertical coordinate of the top of the bedload layer, $z_s$, is defined as that where $\left< \phi \right>$ = $1/2 \left< \phi_{sat} \right>$, and  the vertical coordinate $z_c$ where the transition from creep to bedload occurs is determined by the the kink in $\left< V_x \right>$ (the values of $z_c$ are shown in Fig. \ref{fig:profiles}d).

\subsection{Grain segregation}

\begin{figure}[h!]
	\begin{center}
		\includegraphics[width=.85\linewidth]{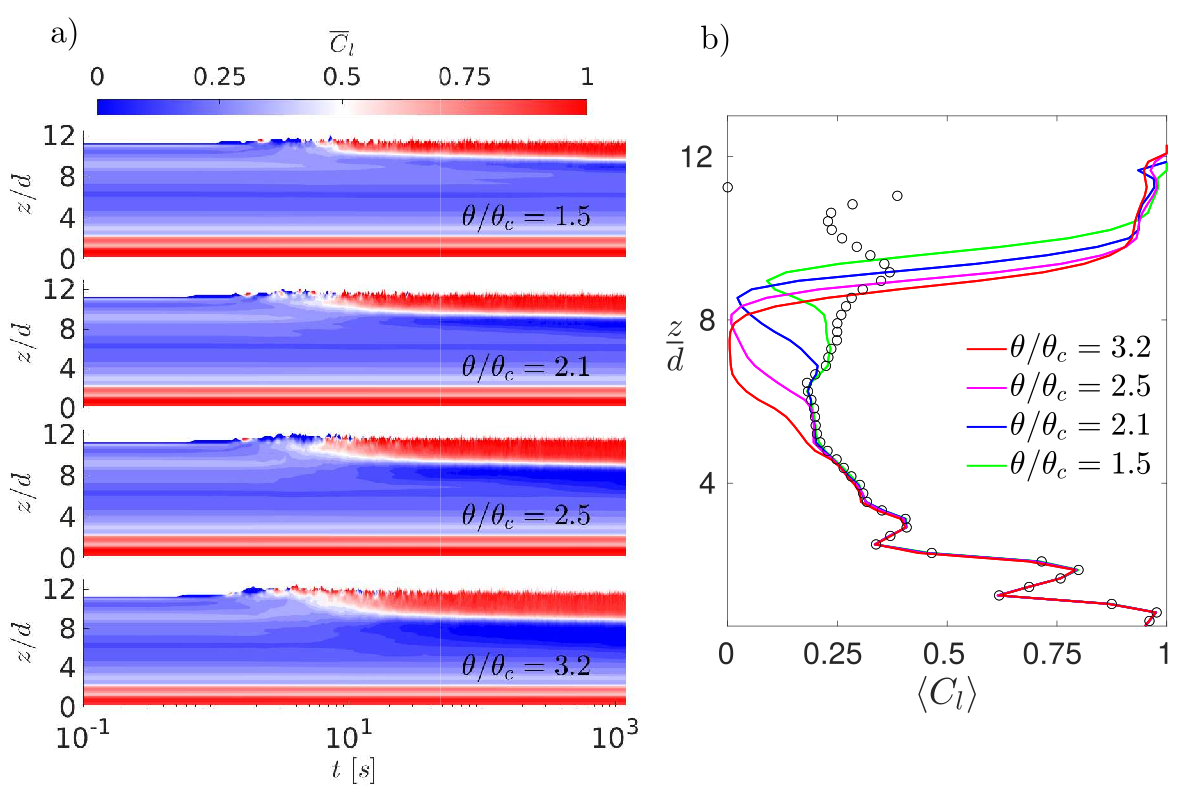}\\
	\end{center}
	\caption{Concentration of large particles $\overline{C}_l$ as a function of the vertical position $z$ normalized by $d$. (a)  Time evolution of $\overline{C}_l(z)$ (space-time diagram). (b) Profiles of $\left< C_l \right>$ within the last 200 s (continuous lines), and for the beginning of simulations (circles).}
	\label{fig:segregation_profiles}
\end{figure}

Figure \ref{fig:segregation_profiles}a shows a space-time diagram of the space-averaged ($x$ and $y$ directions) concentration of large particles $\overline{C}_l$, for all stresses used in the simulations. We observe that segregation takes less time and is stronger as the imposed stress increases, beginning at $t$ $\sim$ 10 s when $\theta / \theta_c$ = 1.5 and at $t$ $\sim$ 1 s when $\theta / \theta_c$ = 3.2. Figure \ref{fig:segregation_profiles}b shows profiles of $\left< C_l \right>$ within the last 200 s (space-time averages), and at the beginning of simulations. From this graphic, it is clear that by the end of simulations $\left< C_l \right>$ on the top of the bed is higher for higher values of $\theta / \theta_c$. This segregation leads to bed armoring.

\begin{figure}[h!]
	\begin{center}
		\includegraphics[width=.45\linewidth]{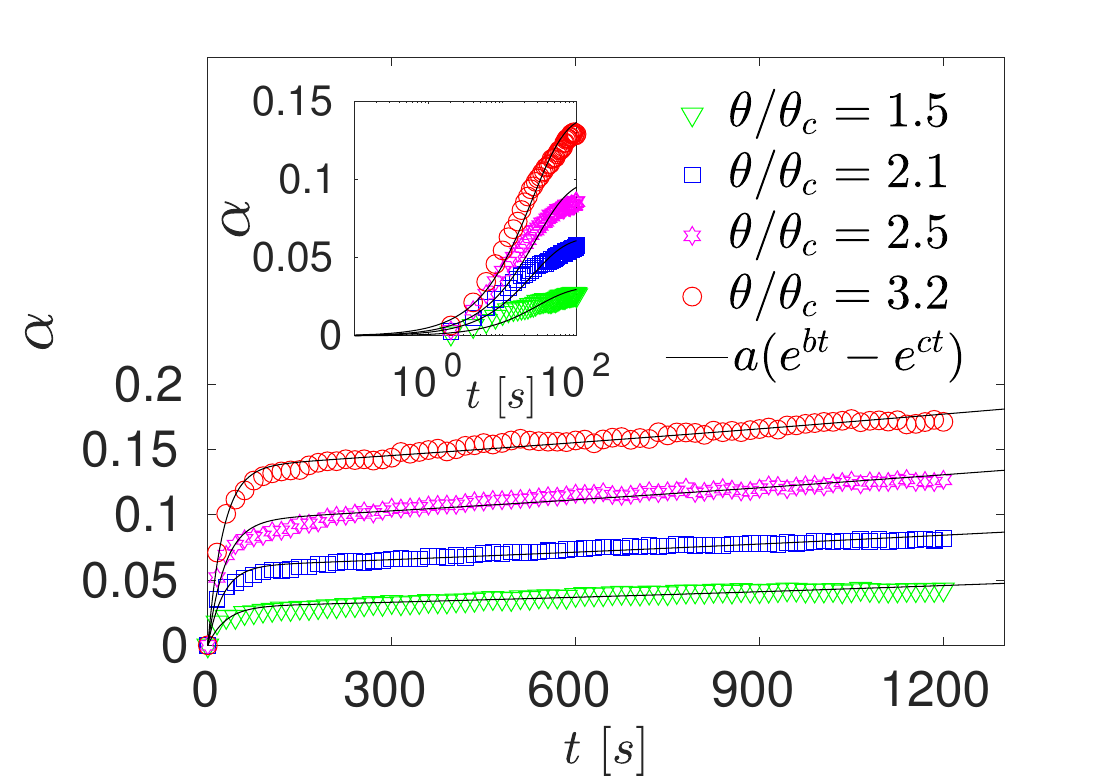}\\
	\end{center}
	\caption{Time evolution of the degree of segregation $\alpha$ for the different shear stresses applied. The symbols are listed in the key and the solid line corresponds to an exponential fit. The inset shows a zoom in the region close to $t$ = 0 s.}
	\label{fig:alpha}
\end{figure}

In order to quantify the degree of segregation, $\alpha$, we proceed as in Jing et al. \cite{Jing}:

\begin{equation}
	\alpha = \frac{1}{2} \left( 1 - \frac{\left( z_{small} - z_{large} \right)_t }{\left( z_{small} - z_{large} \right)_{t_0} } \right) \,\,,
\end{equation}

\noindent where $z_{small}$ is the vertical position of the center of mass of small grains, $z_{large}$ is that of large grains, and $t$ and $t_0$ refer to the current and initial times, respectively. Therefore, $\alpha$ = 0 corresponds to the initial condition and $\alpha$ = 1 to a perfect segregation. Figure \ref{fig:alpha} presents the time evolution of $\alpha$ for the different shear stresses used. We observe that, indeed, segregation is stronger for higher shear stresses, and that it is faster during the first few minutes, but keeps occurring until the end of simulations. These results are also in agreement with Gonzalez et al. \cite{Gonzalez}.

\subsubsection{Motion of particles}

\begin{figure}[h!]
	\begin{center}
		\includegraphics[width=.75\linewidth]{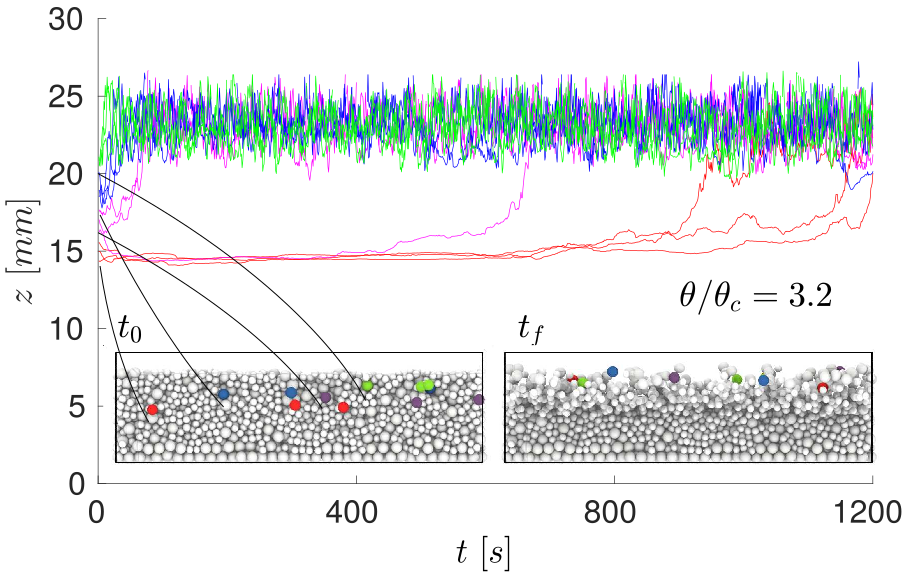}\\
	\end{center}
	\caption{Time evolution of the vertical position $z$ of some large particles moving upward, for $\theta / \theta_c$ $=$ 3.2. We tracked three particles with starting points in the top region of the creep layer, colored in red, and nine particles with starting points in the bedload layer, colored in violet, blue and green as their initial positions are higher within the bed. In the bottom of the figure, the inset on the left shows their respective initial positions, and that on the right their final positions.}
	\label{fig:vertical_motion}
\end{figure}

We tracked the motion of large particles, and analyze next those that moved upward. Figure \ref{fig:vertical_motion} shows the time evolution of the vertical position $z$ of some of those large particles that were initially at different regions within the bed: three particles with starting points in the top region of the creep layer, colored in red, and nine particles with starting points in the bedload layer, colored in violet, blue and green as their initial positions are higher within the bed, as indicated in the insets of Fig. \ref{fig:vertical_motion} (the left inset corresponds to their initial positions and the right one to their final positions). A figure for the upward motion of large particles with starting point at the creep-bedload transition, for the different shear stresses simulated, is available in the supplementary material. As in the experiments of Gonzalez et al. \cite{Gonzalez}, particles in upper layers segregate faster than those in other layers, and this behavior is observed for all segregating grains, as shown in the supplementary material. However, different from Gonzalez et al. \cite{Gonzalez}, we inquire next into the relative roles of contacts and direct fluid entrainment in segregation in the different regions \corr{(a diagram illustrating the problem is available in the supplementary material)}.

\subsubsection{Contact network}
\label{subsection_contact_network}

\begin{figure}[h!]
	\begin{center}
		\includegraphics[width=.99\linewidth]{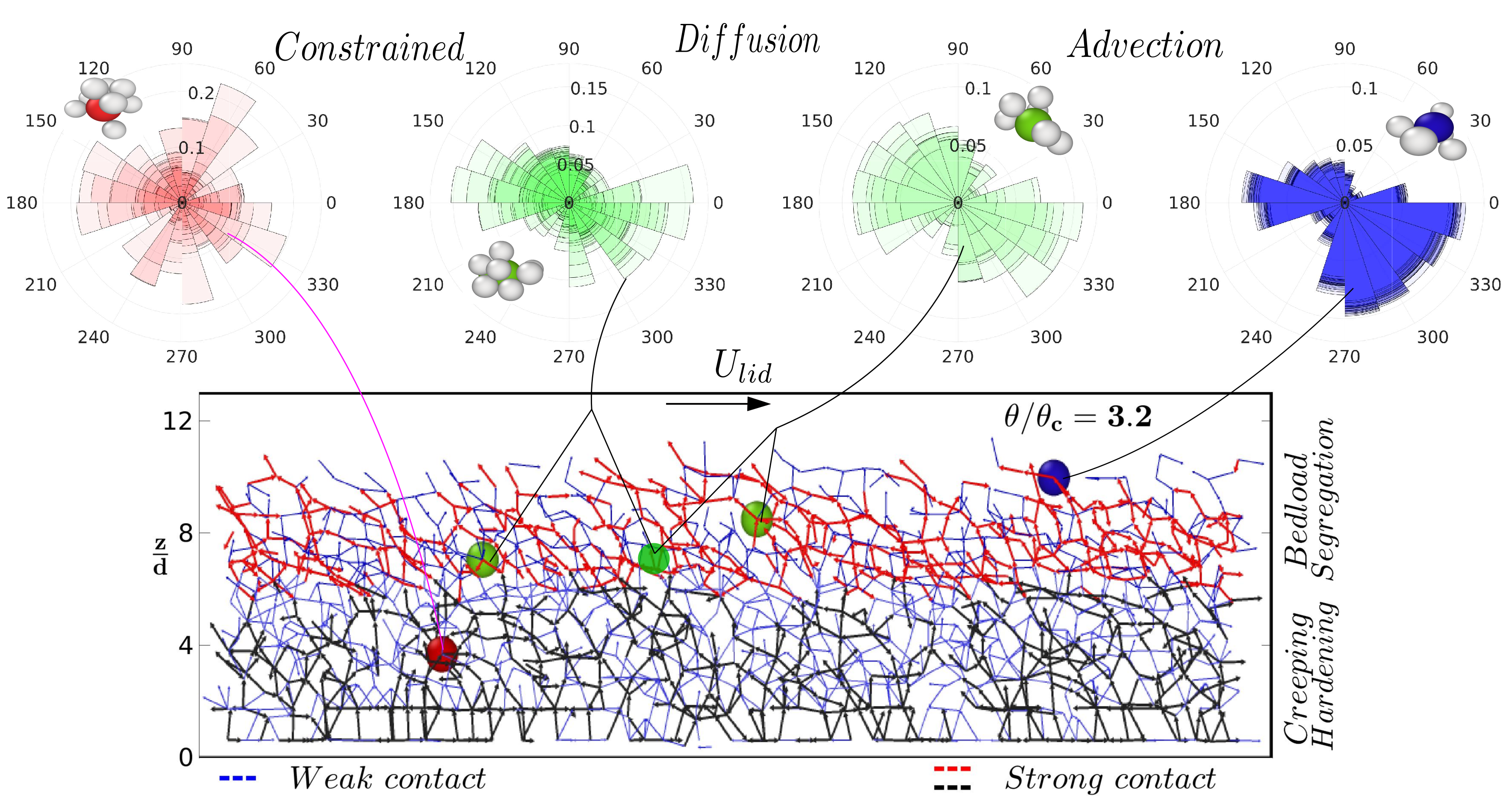}\\
	\end{center}
	\caption{On the bottom, contact network at a given instant, with some particles marked in three different regions:  middle of the creep layer (red particle), bottom and middle of the bedload layer (in green), and top of the bedload layer (in blue). On the top, frequencies of occurrence of contacts as functions of the angular position around the particle, computed for all the large grains in the indicated regions \corr{(Probability density functions -- PDFs -- of these contacts are available in the supplementary material)}. Multimedia available online.}
	\label{fig:contacts}
\end{figure}

Figure \ref{fig:contacts} (Multimedia view, a high-resolution movie also available in an open repository \cite{Supplemental}) shows the contact network at a given instant, and the frequencies of occurrence of contacts as functions of the angle around the particle, computed for all the large grains in the indicated regions (we marked some particles for illustration purposes). These regions correspond to the middle of the creep layer (red particle), bottom and middle of the bedload layer (two particles in green, for which we show the downstream one in two different positions), and top of the bedload layer (surface, particle in blue). Because simulations were 3D and the plot is 2D, the figure on the bottom represents a cut in the bed (see the supplementary material for a figure illustrating how the plane was chosen), and the polar graphics in the figure correspond to the projections of the contacts on the $y$ plane. We observe that contacts in the creep layer occur roughly in all directions, but with peaks within 30$^\circ$ and 90$^\circ$, within 150$^\circ$ and 200$^\circ$, within 210 $^\circ$ and 280$^\circ$, and within 300$^\circ$ and 0$^\circ$. In addition, as showed in Gonzalez et al. \cite{Gonzalez}, the motion of large particles in this region is constrained.

In the bedload layer, the frequencies of occurrence of contacts are different according to the region the large particles are in. At the bottom of the bedload layer (represented by the two green particles at lower $z$ values), contacts become anisotropic, with large peaks within 150$^\circ$ and 200$^\circ$ (upstream face of particles) and within 310$^\circ$ and 20$^\circ$ (median value equal to -15$^\circ$, at the downstream face of particles). Therefore, the large grains in this region are mainly pushed in the longitudinal direction, with a small vertical component given by the moment around an average -15$^\circ$ contact. This can be seen in the movie (multimedia view) associated with this figure (the initial motion of the green particles). Large particles in this region have a diffusion-like vertical motion, as shown by Ref. \cite{Gonzalez}. For the middle of the bedload layer, the distribution of contacts becomes mode inclinated with respect to the horizontal, most of them occurring within 90$^\circ$ and 210$^\circ$, and within 270$^\circ$ and 0$^\circ$ (median value equal to -45$^\circ$). The large grains in this region experience a faster upward motion, mainly due to the moment around the -45$^\circ$ contact. This can also be seen in the movie (multimedia view, the final motion of one of the green particles), and Ref. \cite{Gonzalez} showed that large particles in this region have an advection-like vertical motion. These particles eventually reach the bed surface (represented by the blue particle), and most of the contacts remain only within 270$^\circ$ and 0$^\circ$ (median value equal to -45$^\circ$), in a balance with direct fluid entrainement (fluid forces).

The analysis of the frequency of occurrence of contacts as a function of the orientation leads us to infer that in the top of the creep layer and bottom of the bedload layer, the upward motion of large grains is driven mainly by solid-solid contacts, while in the middle and top of the bedload layer it is driven by direct fluid entrainment together with solid-solid contacts. We investigate this in detail in Subsection \ref{subsection_forces}, and find that this hypothesis is, indeed, true. 

\begin{figure}[h!]
	\begin{center}
		\includegraphics[width=.99\linewidth]{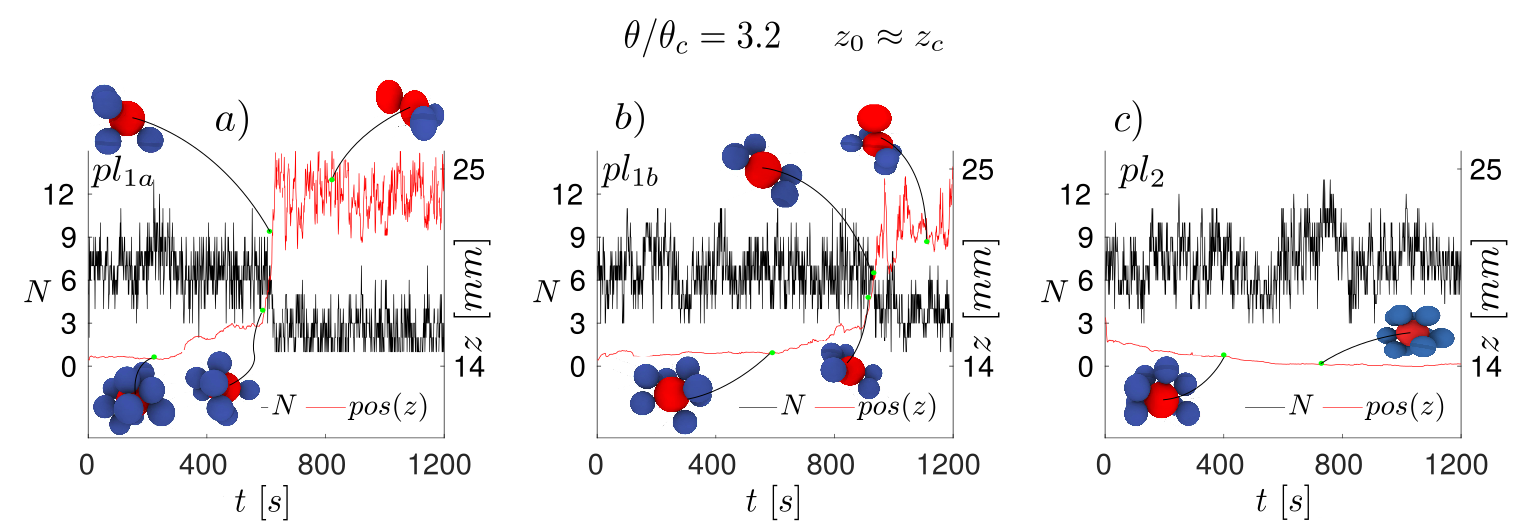}\\
	\end{center}
	\caption{Evolution of the number of contacts $N$ (black lines) and of the vertical position $z$ (red lines) along time for (a)-(b) some large grains in the middle and bottom of the bedload layer, moving upward, and (c) a large grain, not segregating, in the bottom of the bedload layer. The panels present also the large particle (in red) at some instants identified in the graphics, and the grains in contact with it (the blue ones are the small grains).}
	\label{fig:contacts2}
\end{figure}

The contact network can also be evaluated in terms of the number of contacts $N$ of particles along time. For example. we selected some large particles and computed $N$ along time, which we show in Fig. \ref{fig:contacts2} together with their vertical positions $z$, for $\theta / \theta_c$ = 3.2. Figure \ref{fig:contacts2} also shows snapshots of contacts during the upward motion of large grains, at some instants identified in the graphics, with large particles being identified in red and small particles in blue. With that, we can observe variations in both $N$ and orientation of contacts as the large particle moves upward. We observe that, indeed, $N$ decreases when the particle moves upward (seen in Figs. \ref{fig:contacts2}a and \ref{fig:contacts2}b, and absent in Fig. \ref{fig:contacts2}c), going from roughly 6 contacts to approximately 3 contacts. This is, in a certain way, in agreement with the formation of a contact network oriented in the -45$^\circ$ direction as time evolves. We observe that the contacts in the downstream direction at the instant of the upward motion, shown in Figs. \ref{fig:contacts2}a and \ref{fig:contacts2}b, are, indeed, aligned in -45$^\circ$. This indicates that contact and/or fluid forces push the large particle in the flow direction, which then turns over the -45$^\circ$ contact point (due to the moment around this point). Therefore, in the upward motion from the creep to the bedload layer, and throughout the bedload layer, solid-solid contacts are important, with a moment about a -45$^\circ$ contact point with respect to the horizontal plane. Figure \ref{fig:contacts2}c shows a large particle that, although in the bottom of the bedload layer, does not segregate (absence of upward motion). Instead, it remains trapped and moves downward together with the small particles in this region due to the increase in bed compaction. We note that this large particle maintains (in average) the number of contacts constant (with a mean value around 6-7), and contacts aligned in the -45$^\circ$ direction are not observed.

\subsubsection{Forces on particles}
\label{subsection_forces}

\begin{figure}[h!]
	\begin{center}
		\includegraphics[width=.99\linewidth]{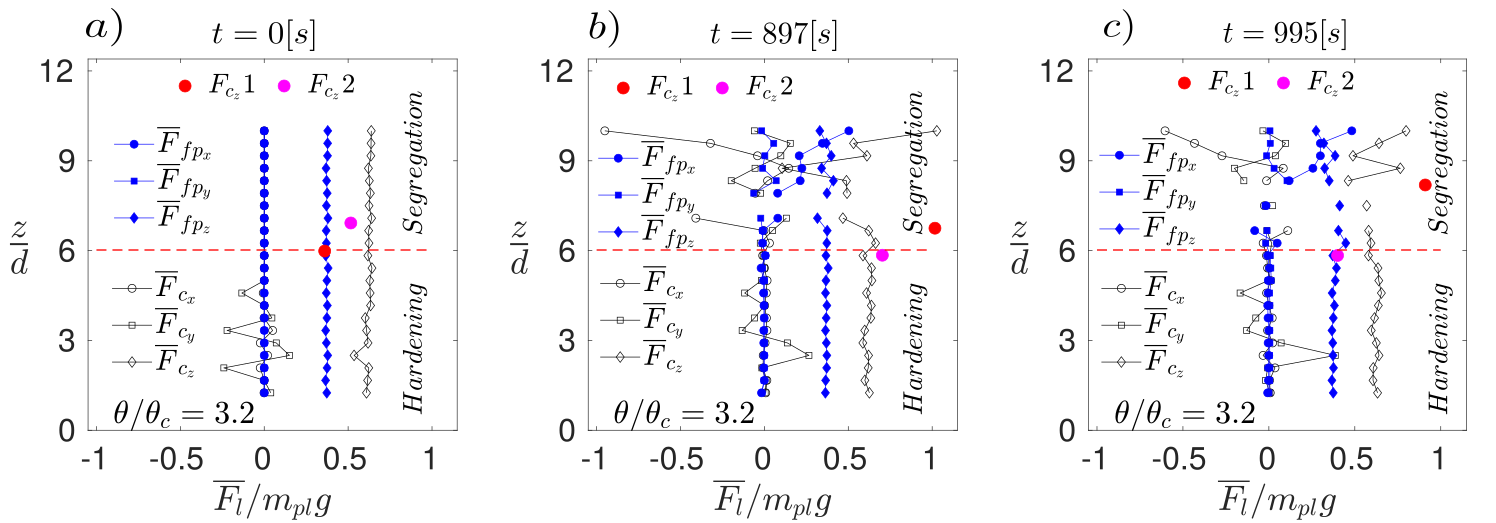}\\
	\end{center}
	\caption{Vertical profiles of space-averaged contact and fluid forces acting on large grains, $\overline{F}_l$, \corr{normalized by the weight of a large grain, $m_{pl}g$,} at three different instants: (a) $t$ = 0 s; (b) $t$ = 897 s; (c) $t$ = 995 s. The solid symbols correspond to the components of the resultant force caused by the fluid on the particle and the open symbols to that caused by the contacts, the symbols being listed in the key. The figure shows the vertical positions (ordinate) and vertical component of the contact force ($F_{c_z}$, read in the abscissa) of two large particles: one in red, which moves upward along time, and another one in magenta, which remains trapped and moves downward due to an increase in bed compaction. Space averages are computed in the longitudinal and spanwise directions, by considering the large grains only, and $\theta / \theta_c$ = 3.2.}
	\label{fig:forces_proportions}
\end{figure}

Because in the CFD-DEM simulations we compute Newton's second law for each solid particle at each time step, we have access to each force listed in Eqs. \ref{Fp2} and \ref{Ffp_sim} as well as the resultant force $\vec{F}_{p}$. This is an information that is not accessible in experiments or field measurements, and is important for understanding the motion of grains leading to their reorganization within the bed. In this section, we inquire into the roles of the forces that are either due directly to the fluid or transmitted exclusively through solid-solid contacts, for the vertical motion of large grains

Figures \ref{fig:forces_proportions}a-c show the vertical profiles of space-averaged components of the resultant contact and fluid forces acting on large grains, $\overline{F}_l$, \corr{normalized by the weight of a large grain, $m_{pl}g$,} at $t$ = 0,  897, and 995 s, respectively \corr{(where $m_{pl}$ is the mass of a large particle)}. The symbols are listed in the figure key, and the figures show the positions and vertical component of contact forces of two large particles, the red symbols corresponding to a particle that moves upward along time, and the magenta one to a particle that remains trapped and moves downward due to the increase in bed compaction (they are tracked along the three panels). \corr{A dimensional form of Fig. \ref{fig:forces_proportions} is available in the supplementary material.} 

At $t$ = 0 s, the longitudinal and transverse components of forces caused by the fluid ($F_{fp_{x}}$ and $F_{fp_{y}}$, respectively) and solid-solid contacts ($F_{c_{x}}$ and $F_{c_{y}}$, respectively) are zero, since the flow has not yet started, while the vertical components of fluid $F_{fp_{z}}$ and contact $F_{c_{z}}$ forces correspond to \corr{buoyancy ($\approx$ 0.4 $m_{pl}g$, which means 1.3 $\times$ 10$^{-4}$ N) and load distribution ($\approx$ 0.6 $m_{pl}g$, which means values within} 2.1 $\times$ 10$^{-4}$ and 2.2 $\times$ 10$^{-4}$ N), respectively. Due to the absence of flow, the profiles are roughly uniforms, with some small fluctuation in the contact forces caused by the initialization of simulations (settling by gravity). At $t$ = 897 s, after the liquid flow has been imposed, we observe that in the creep region (identified as hardening) the profiles of all components of $F_{fp}$ and $F_{c}$ remain approximately uniform\corr{, being} roughly the same as \corr{those} at $t$ = 0 s. In the bedload region (identified as segregation), the profiles deviate considerably from their uniform values. We observe that $F_{fp_{x}}$ increases from zero at the creep-bedload transition ($z$ = $z_c$) to \corr{$F_{fp_{x}}/(m_{pl}g)$ $\approx$ 0.6 ($F_{fp_{x}}$ $\approx$ 2 $\times$ 10$^{-4}$ N)} close to the bed surface. \corr{This maximum value is approximately equal to the modulus of the relative weight $W_{rel}$ of large particles ($W_{rel}$ = (1 $-$ 0.37) $m_{pl}g$), meaning that $F_{fp_{x}}$ can directly entrain the grains that are at the bed surface. Therefore, $F_{fp_{x}}$ is the driving force, and} the average contact force in the longitudinal direction, $F_{c_{x}}$, is a reaction to $F_{fp_{x}}$ whenever grains are in contact, \corr{increasing from} zero at $z$ = $z_c$ to \corr{$F_{c_{x}}/(m_{pl}g)$ $\approx$ $-$0.9 ($F_{c_{x}}$ $\approx$ $-$3 $\times$ 10$^{-4}$ N)}. Because the contact chains are rather complex, the $F_{c_{x}}$ profile has higher fluctuations when compared \corr{with $F_{fp_{x}}$}. Concerning the vertical components, $F_{fp_{z}}$ oscillates around the value found at the creep layer (that are due to buoyancy), decreasing for the topmost grain (probably due to flow accelerations experienced by the exposed grains), while the $F_{c_{z}}$ profile increases in the $z$ $\gtrsim$ 9$d$ region due to \corr{-45$^\circ$ contact points: the particles are pushed in the longitudinal direction by the direct action of the fluid or via longitudinal contacts, and roll over a -45$^\circ$ contact point} (as shown in Subsection \ref{subsection_contact_network}).

Between $t$ = 0 s and $t$ = 897 s, the particle indicated in red moved upward of a distance corresponding to one grain diameter, while the particle indicated in pink moved downward a distance also of one grain diameter. At $t$ = 995 s, the force profiles are roughly similar to those at $t$ = 897 s\corr{: the main} behavior of profiles, including the order of magnitude of forces, remains approximately the same, \corr{we notice only small} variations in the fluctuations present in the profiles. Between $t$ = 897 s and $t$ = 995 s, the particle indicated in pink remained trapped in the same vertical position, while the one in red moved upward a distance corresponding to approximately one grain diameter. 
 
\begin{figure}[h!]
	\begin{center}
		\includegraphics[width=.7\linewidth]{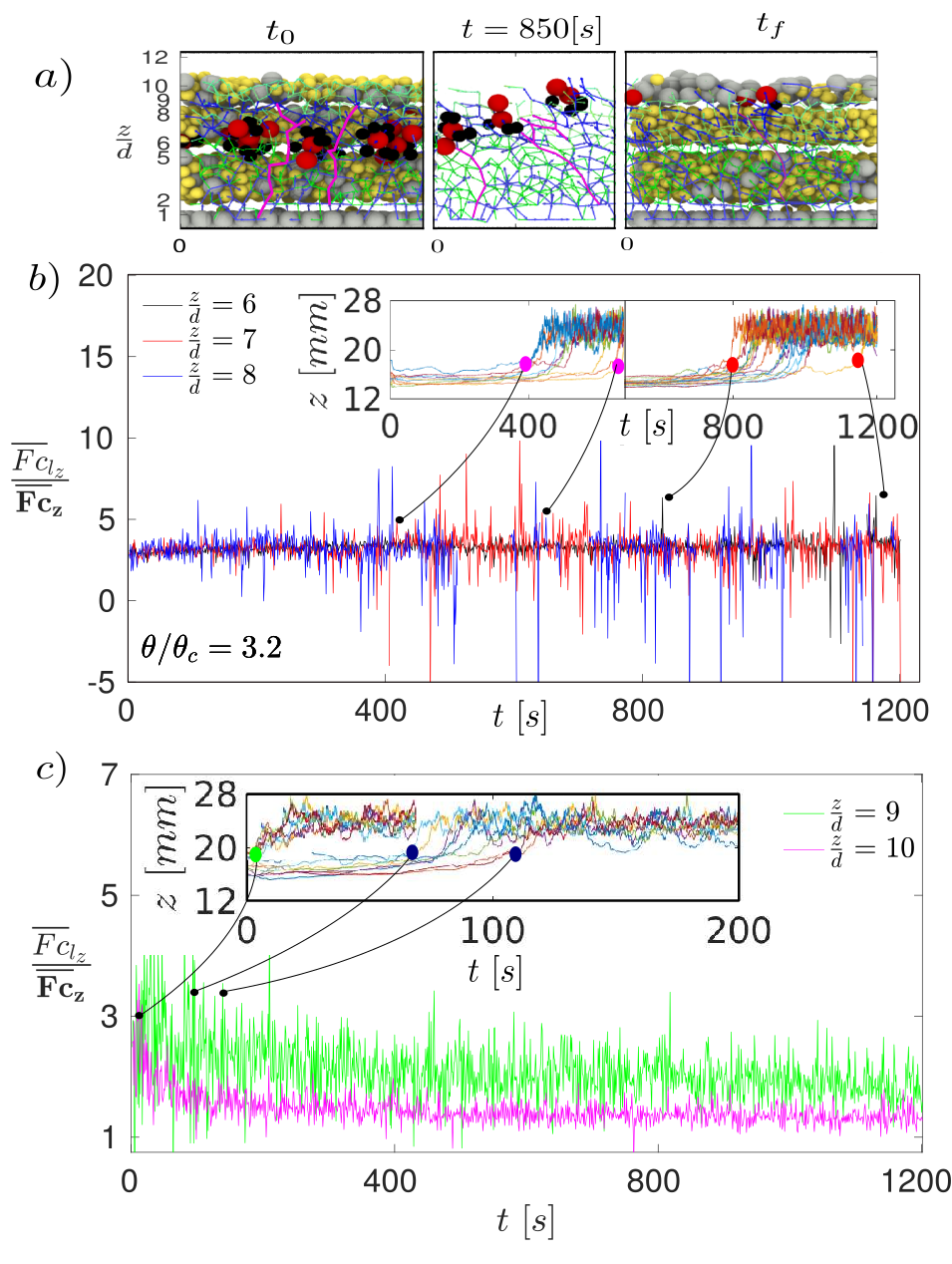}\\
	\end{center}
	\caption{Contact forces on large grains at different heights within the bed. (a) Positions of some of the tracked grains at three different instants: $t$ = $t_0$ (initial), $t$ = 850 s, and $t$ = $t_f$ (final). \corr{Some particles were removed for visualization purposes, so that empty layers appear in the figures (these layers are not really present in the simulations).} (b) Space average of the vertical component of contact forces for large grains at different heights, for $z/d$ from 6 to 8, normalized by the average for each height (including the small grains), $\overline{F}_{cl_{z}} / \overline{F}_{c_{z}}$. (c) $\overline{F}_{cl_{z}} / \overline{F}_{c_{z}}$ for $z/d$ from 9 to 10. The insets of panels (b) and (c) show the instantaneous positions of all large grains for $z/d$ between 6 and 8 and between 9 and 10, respectively, with some upward motions indicated in the main graphics.}
	\label{fig:forces_segregation}
\end{figure}

In order to evaluate the magnitude of the contact forces at different heights within the bed, we computed the space averages of the vertical component of contact forces acting on large grains, $\overline{F}_{cl_{z}}$, at each time step and different heights. The space averages were computed only in the $x$ and $y$ directions. Figures \ref{fig:forces_segregation}b and \ref{fig:forces_segregation}c show $\overline{F}_{cl_{z}}$ normalized by the average \corr{force} for each height (including the small grains), $\overline{F}_{c_{z}}$, for $z/d$ from 6 to 8 and from 9 to 10, respectively (a figure showing the contact forces for the small grains is available in the supplementary material). Examples of the relative positions of large particles within the bed are shown in Fig. \ref{fig:forces_segregation}a. \corr{We note that some particles were removed in Fig. \ref{fig:forces_segregation}a for visualization purposes, so that empty layers appear in this figure (these layers are not really present in the simulations).}

The inset in Fig. \ref{fig:forces_segregation}b shows the instantaneous positions of all large grains within $z/d$ between 6 and 8, with some of the upward motions indicated in the main graphic. We observe that in this region the upward motions result from a positive peak in $\overline{F}_{cl_{z}}$, followed by negative values when the particle stops the vertical displacement. In general, the amplitudes of positive peaks reach values within 5 and 10, higher than those found in the region for $z/d$ between 9 and 10 (Fig. \ref{fig:forces_segregation}c), where positive peaks reach values within 3 and 4. This corroborates the observations made in Subsection \ref{subsection_contact_network}, that the vertical contacts are particularly important for the upward motion of large grains in the creep-bedload transition and in the lower part of the bedload layer. In the upper part of the bedload layer, fluid forces entrain the large grains longitudinally, which migrate upwards via contacts aligned in the -45$^\circ$ direction. 

\begin{figure}[h!]
	\begin{center}
		\includegraphics[width=.99\linewidth]{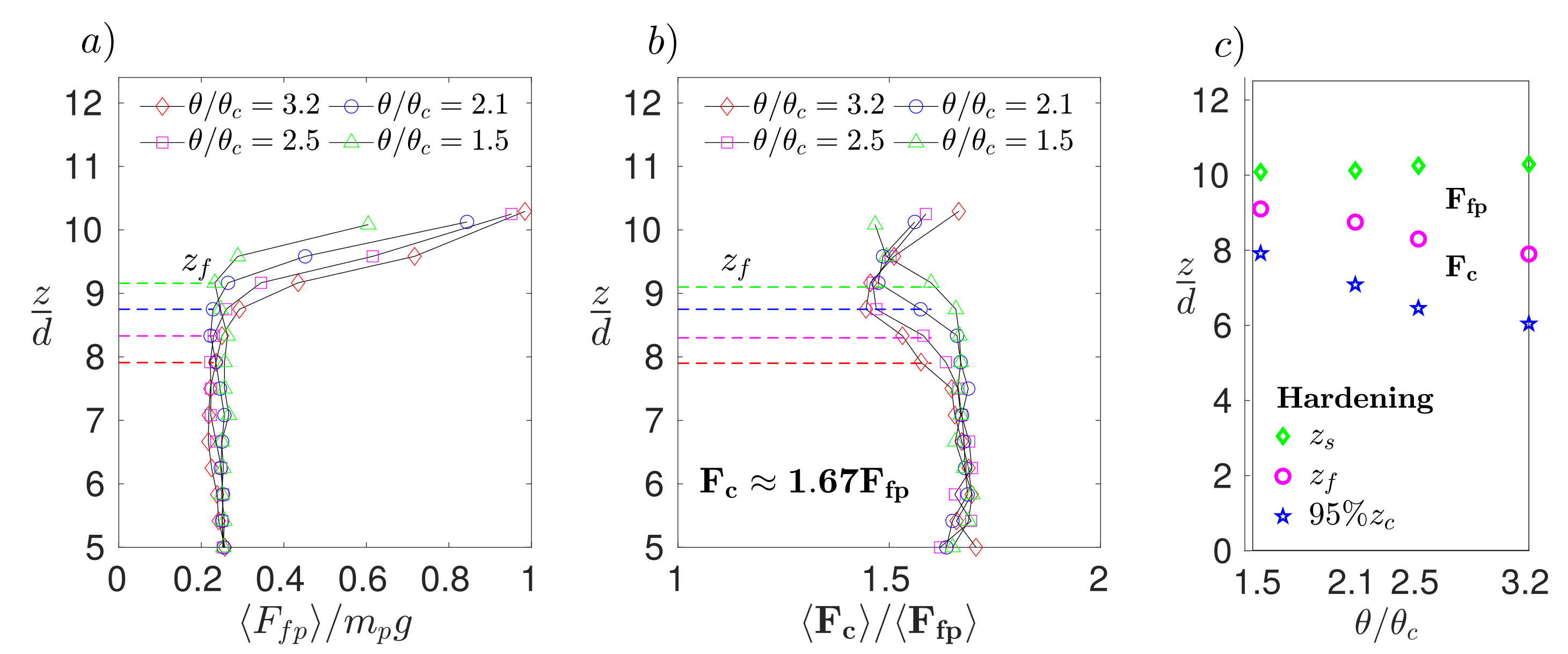}\\
	\end{center}
	\caption{(a) Vertical profiles of the space-time average of the forces caused by the fluid on all grains, $\left< F_{fp} \right>$, \corr{normalized by an average weight of grains $m_pg$ based on the volume-averaged diameter $d$,} for different values of $\theta / \theta_c$. (b) vertical profiles of the space-time averages of contact forces, $\left< F_{c} \right>$, normalized by $\left< F_{fp} \right>$, for different values of $\theta / \theta_c$. (c) Heights of the bedload surface, $z_s$, the top 95\% of the creep layer, 0.95$z_c$, and where $\left< F_{fp} \right>$ deviates from values found in the creep layer, $z_f$.}
	\label{fig:forces_proportions2}
\end{figure}

Finally, to find the heights within the bed where fluid forces affect directly the motion of particles, we computed \corr{space-time averages} of the forces caused by the fluid on grains (all of them, both large and small grains) $\left< F_{fp} \right>$\corr{. The profiles of $\left< F_{fp} \right> / m_pg$, where $m_pg$ is an average weight of grains based on the volume-averaged diameter $d$, are shown in Fig. \ref{fig:forces_proportions2}a (a dimensional form of this graphic is available in the supplementary material)}. We observe that values increase monotonically from \corr{$\approx$ 0.25 $m_pg$ (0.45 $\times$ 10$^{-4}$ N)} within the creep layer, which corresponds to the average buoyancy force, to values within \corr{$\approx$ 0.6 and 1.0 $m_pg$ (1.0 $\times$ 10$^{-4}$ and 1.7 $\times$ 10$^{-4}$ N)} on the bed surface. We also indicate in the figure the heights where $\left< F_{fp} \right>$ start deviating from the buoyancy force, which we call $z_f$, the values of $z_f$ varying within 8 and 9 (Fig. \ref{fig:forces_proportions2}c). Below $z_f$, which corresponds to the lower part of the bedload layer and the creep layer, the forces are transmitted exclusively by solid-solid contacts (fluid forces do not act directly on the motion of large grains). Above $z_f$, which corresponds to the middle and upper regions of the bedload layer, the direct action of fluid forces \corr{becomes} significant for the motion of large grains: the fluid thrust the grains, and the contacts at -45$^{\circ}$ make the grains move upward. \corr{In terms of imposed stresses, we observe that not only the magnitude of fluid forces becomes higher as the stress increases, but also the region where the fluid forces (other than buoyancy) act directly on particles increases (i.e., $z_f$ decreases with increasing $\theta$).}

Figure \ref{fig:forces_proportions2}b shows vertical profiles of the space-time averages of contact forces $\left< F_{c} \right>$ normalized by $\left< F_{fp} \right>$, and we observe that in the creep layer and in the bottom of the bedload layer (i.e., below $z_f$), $\left< F_{c} \right> / \left< F_{fp} \right>$ is roughly 1.7. This happens because the fluid accounts only for the buoyancy force in this region, and the contacts transmit only the relative weight of particles, so that $\left< F_{c} \right> / \left< F_{fp} \right>$ $\approx$ $S$ $-$ 1 = 1.7. For $z$ $>$ $z_f$, in the region where fluid forces other than buoyancy affect directly the grains, $\left< F_{c} \right> / \left< F_{fp} \right>$ decreases, reaching values of approximately 1.5. As a consequence of the direct action of the fluid on grains, this region dilates \cite{Cunez2} and the average number of contacts decrease (as shown in Fig. \ref{fig:anisotropy} of Subsection \ref{subsection_anisotropy}). For the topmost grains, at $z$ = $z_s$, $\left< F_{c} \right> / \left< F_{fp} \right>$ increases again, with both $\left< F_{c} \right>$ and $\left< F_{fp} \right>$ increasing: the space-average contact force on those grains increases because they have only contacts on their bottom, producing higher resultant of contact forces. Therefore, the analysis of forces acting on large grains is in agreement with the observation made based on solid-solid contacts in Subsection \ref{subsection_contact_network}: (i) in the top and middle of the bedload layer ($z$ $\geq$ $z_f$), fluid forces thrust grains in the main flow direction, and the reaction on contact points around -45$^{\circ}$ (contact forces) makes them move upward; (ii) in the bottom of the bedload layer and in the creep layer ($z$ $<$ $z_f$), the particles are not thrust directly by the fluid, but by the transmission of forces via contact chains, so that the upward motion of grains in this region is driven by contacts only.

\corr{We note that our findings are in agreement with those of Ferdowsi et al. \cite{Ferdowsi}, although in their experiments $S$ $=$ 1.13 (close to unity). Our results are also in agreement with Rousseau et al. \cite{Rousseau}, who measured the upward motion of one larger particle (intruder) amid smaller grains undergoing bedload motion. Based on exhaustive experiments, they found that the upward motion is initially (deeper in the bed) intermittent and slow, and after a while (closer to the bed surface) it becomes faster. Later, Gonzalez et al. \cite{Gonzalez} inferred, based on their experimental results, that the initial phase of the intruder's motion would correspond to the limit between creep and bedload. The present results indicate that, in fact, the initial phase corresponds not only to the limit between creep and bedload, but also to the lower part of the bedload layer. In these regions, forces pushing the grains are exclusively transmitted via solid-solid contacts, while in the middle and top of the bedload layer fluid forces also act directly on particles. In addition, we found that in both cases the large particle is pushed in the longitudinal direction and roll over a -45$^{\circ}$ contact point.}

\corr{We note also that, depending on the point of view, conclusions about weakening of segregation due to the presence of a viscous fluid must be taken with care. We found that, under the action of a shearing fluid, particles in the middle and upper parts of bedload are also pushed directly by the fluid (in addition to solid-solid contacts), so that they roll over other particles. In this sense, the fluid flow helps segregation in those regions. However, if we compare the motion of grains with and without the presence of a fluid (not the objective here), viscous effects dampen segregation, as shown by  Zhou et al. \cite{Zhou2} and Cui et al. \cite{Cui}.}

\subsection{Bed hardening}
\label{subsection_anisotropy}

\begin{figure}[h!]
	\begin{center}
		\includegraphics[width=.9\linewidth]{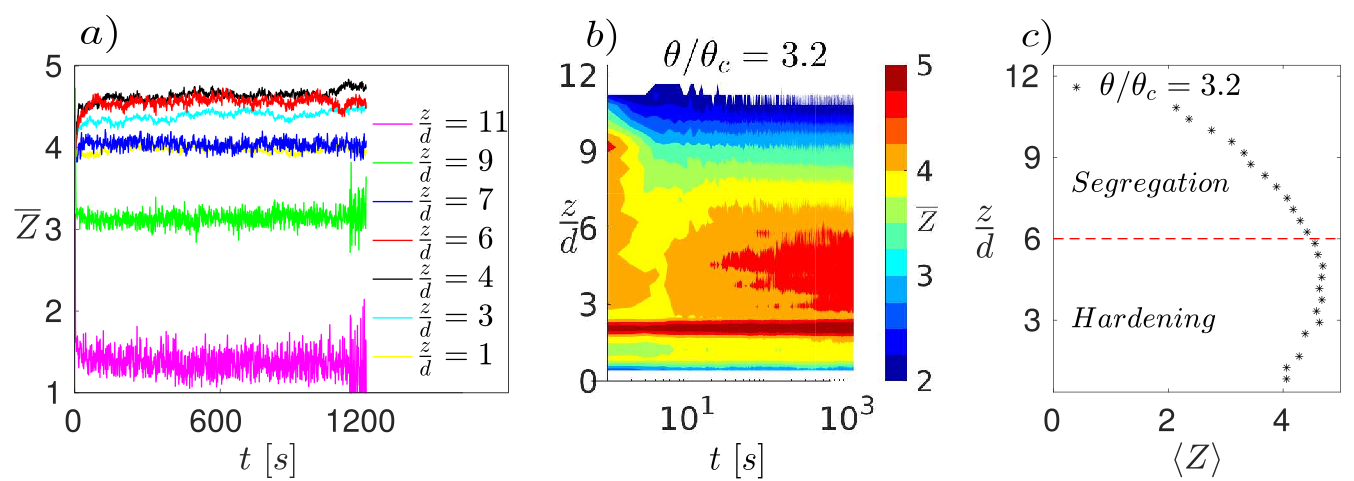}\\
	\end{center}
	\caption{(a) For different heights, the space averages ($x$-$y$ average for the considered height) of the coordination number, $\overline{Z}$ as a function of time. (b) Space-time diagram of $\overline{Z}$. (c) Vertical profile of the space-time average of the coordination number, $\left< Z \right>$.}
	\label{fig:anisotropy}
\end{figure}

Finally, we investigate the time evolution of the solid-solid contacts, and their effects on bed hardening. For that, we computed the coordination number of particles (average number of contacts per grain) averaged in the $x$ and $y$ directions, $\overline{Z}$, and also the space-time average of the coordination number, $\left< Z \right>$. Figure \ref{fig:anisotropy}a shows $\overline{Z}$, computed at different heights, as a function of time. We observe that at higher heights ($z/d$ = $9$) values are smaller (within 3 and 4), since those grains are close to the bed surface (the contacts occur only on their bottom). In the creep and bedload layers, values vary within 4 and 5 without a clear dependence on $z$ from this figure, but increasing during the first seconds to reach values roughly constant afterward. In order to analyze the variation of the coordination number along time, Fig. \ref{fig:anisotropy}b shows the space-time diagram of $\overline{Z}$. We observe a significant decrease in the average number of contacts close to the bed surface ($z/d$ between 9 and 12), $\overline{Z}$ decreasing over time from approximately 4 to approximately 2. Below this region, for $z/d$ within 8 and 9 (middle of the bedload layer), $\overline{Z}$ decreases more smoothly over time, from approximately 4 to approximately 3.5, while for $z/d$ within 6 and 7 (bottom of the bedload layer) the number of contacts remains roughly constant, decreasing and then increasing ($\overline{Z}$ remaining very close to 4 at all times). The decrease in the top of the bedload layer and the constant value in its bottom is consistent with the dilation nature of bedload \cite{Cunez2}. In contrast, for $z/d$ within 2.5 and 6 (creep layer), the average number of contacts increases, $\overline{Z}$ going roughly from 4 to 5, which is consistent with compaction and hardening in this region \cite{Cunez2}. The time evolution of average contacts in the bed is corroborated in terms of space-time averages, shown in Fig. \ref{fig:anisotropy}c. \corr{Similar graphics for either large or small particles are available in the supplementary material. In terms of space-time diagrams and space-time averages, the behavior is roughly the same for each species or both combined. The main differences are: (i) that the large particles have in average more contacts than the small ones; and (ii) that the time evolutions of small and large particles present stronger fluctuations close to the surface and middle of the bedload layer, respectively.}

We also plotted the network of contact forces, that are shown in the supplementary material, and also in Fig. \ref{fig:forces_segregation}a. Those figures show that, as time goes on, load-bearing chains (carrying strong forces \cite{Radjai, Cates}) tend to be aligned in the -45$^{\circ}$ direction, hardening the bed.

\section{CONCLUSIONS}
\label{sec:conclusions}

In this paper, we investigated the dynamics of grains of a bidisperse bed sheared by a viscous liquid by carrying out CFD-DEM simulations, and inquired into the mechanisms leading to size segregation and bed hardening. The numerical results reproduced well the bed dynamics measured in previous experiments \cite{Gonzalez}, with a bedload (fluid-like) layer above a creep (solid-like) layer. The velocity profiles agreed well with those measured experimentally under the same conditions, with a kink between the creep and bedload layers, and an increase of the thickness of the bedload layer with the imposed shear stress. We showed that the upward motion of large particles (segregation) is faster and stronger for larger stresses, and that it is due to contact forces in the creep layer and lower part of the bedload layer. In the middle and upper parts of the bedload layer, the upward motion is due to the direct action of fluid forces combined with transmission via solid-solid contacts. In all cases the particles experience a moment about a -45$^\circ$ contact point with respect to the horizontal plane when  migrating upward, whether entrained by other contacts or directly by the fluid. In addition, we showed the time evolution of the vertical position of large grains (which agrees with those from experiments), the evolution of the average number of contacts, and the increase in bed hardening over time. Finally, we showed the average profiles within the bed of the forces caused either directly by the fluid or transmitted by solid-solid contacts. \corr{We note that the duration of simulations was much smaller than that of previous experiments, so that the bed had not enough time to reach the same level of hardening as in the experiments. We note also that we investigated only a 3:2 size ratio, and the segregation rates can be different for other ratios. Even with that, our} results bring new insights into the mechanisms of segregation and hardening in polydisperse beds, such as happens in river beds and landslides.

\section*{AUTHOR DECLARATIONS}
\noindent \textbf{Conflict of Interest}

The authors have no conflicts to disclose

\section*{SUPPLEMENTARY MATERIAL}
See the supplementary material for supporting information on the numerical model, layout of the numerical setup, and additional graphics.

\section*{DATA AVAILABILITY}
The data that support the findings of this study are openly available in Mendeley Data at http://dx.doi.org/10.17632/992j4h96ky \, \cite{Supplemental} and http://dx.doi.org/10.17632/nzt86s3r8p \, \cite{Supplemental2}.

\begin{acknowledgments}
Erick M. Franklin is grateful to the S\~ao Paulo Research Foundation -- FAPESP (Grant No. 2018/14981-7) and Conselho Nacional de Desenvolvimento Cient\'ifico e Tecnol\'ogico -- CNPq (Grant No. 405512/2022-8) for the financial support provided. Jaime O. Gonzalez would like to thank the Petroleum Department of the Escuela Politécnica Nacional, Quito, Ecuador. The authors thank Fernando David C\'u\~nez for the help with the numerical setup.
\end{acknowledgments}

\bibliography{references}

\begin{thebibliography}{38}%
\makeatletter
\providecommand \@ifxundefined [1]{%
 \@ifx{#1\undefined}
}%
\providecommand \@ifnum [1]{%
 \ifnum #1\expandafter \@firstoftwo
 \else \expandafter \@secondoftwo
 \fi
}%
\providecommand \@ifx [1]{%
 \ifx #1\expandafter \@firstoftwo
 \else \expandafter \@secondoftwo
 \fi
}%
\providecommand \natexlab [1]{#1}%
\providecommand \enquote  [1]{``#1''}%
\providecommand \bibnamefont  [1]{#1}%
\providecommand \bibfnamefont [1]{#1}%
\providecommand \citenamefont [1]{#1}%
\providecommand \href@noop [0]{\@secondoftwo}%
\providecommand \href [0]{\begingroup \@sanitize@url \@href}%
\providecommand \@href[1]{\@@startlink{#1}\@@href}%
\providecommand \@@href[1]{\endgroup#1\@@endlink}%
\providecommand \@sanitize@url [0]{\catcode `\\12\catcode `\$12\catcode
  `\&12\catcode `\#12\catcode `\^12\catcode `\_12\catcode `\%12\relax}%
\providecommand \@@startlink[1]{}%
\providecommand \@@endlink[0]{}%
\providecommand \url  [0]{\begingroup\@sanitize@url \@url }%
\providecommand \@url [1]{\endgroup\@href {#1}{\urlprefix }}%
\providecommand \urlprefix  [0]{URL }%
\providecommand \Eprint [0]{\href }%
\providecommand \doibase [0]{https://doi.org/}%
\providecommand \selectlanguage [0]{\@gobble}%
\providecommand \bibinfo  [0]{\@secondoftwo}%
\providecommand \bibfield  [0]{\@secondoftwo}%
\providecommand \translation [1]{[#1]}%
\providecommand \BibitemOpen [0]{}%
\providecommand \bibitemStop [0]{}%
\providecommand \bibitemNoStop [0]{.\EOS\space}%
\providecommand \EOS [0]{\spacefactor3000\relax}%
\providecommand \BibitemShut  [1]{\csname bibitem#1\endcsname}%
\let\auto@bib@innerbib\@empty
\bibitem [{\citenamefont {Bagnold}(1941)}]{Bagnold_1}%
  \BibitemOpen
  \bibfield  {author} {\bibinfo {author} {\bibfnamefont {R.~A.}\ \bibnamefont
  {Bagnold}},\ }\href@noop {} {\emph {\bibinfo {title} {The Physics of Blown
  Sand and Desert Dunes}}}\ (\bibinfo  {publisher} {Chapman and Hall},\
  \bibinfo {address} {London},\ \bibinfo {year} {1941})\BibitemShut {NoStop}%
\bibitem [{\citenamefont {Raudkivi}(1976)}]{Raudkivi_1}%
  \BibitemOpen
  \bibfield  {author} {\bibinfo {author} {\bibfnamefont {A.~J.}\ \bibnamefont
  {Raudkivi}},\ }\href@noop {} {\emph {\bibinfo {title} {Loose boundary
  hydraulics}}},\ \bibinfo {edition} {1st}\ ed.\ (\bibinfo  {publisher}
  {Pergamon Press},\ \bibinfo {year} {1976})\BibitemShut {NoStop}%
\bibitem [{\citenamefont {Houssais}\ \emph {et~al.}(2015)\citenamefont
  {Houssais}, \citenamefont {Ortiz}, \citenamefont {Durian},\ and\
  \citenamefont {Jerolmack}}]{Houssais_1}%
  \BibitemOpen
  \bibfield  {author} {\bibinfo {author} {\bibfnamefont {M.}~\bibnamefont
  {Houssais}}, \bibinfo {author} {\bibfnamefont {C.~P.}\ \bibnamefont {Ortiz}},
  \bibinfo {author} {\bibfnamefont {D.~J.}\ \bibnamefont {Durian}},\ and\
  \bibinfo {author} {\bibfnamefont {D.~J.}\ \bibnamefont {Jerolmack}},\
  }\bibfield  {title} {\enquote {\bibinfo {title} {Onset of sediment transport
  is a continuous transition driven by fluid shear and granular creep},}\
  }\href@noop {} {\bibfield  {journal} {\bibinfo  {journal} {Nat. Commun.}\
  }\textbf {\bibinfo {volume} {6}} (\bibinfo {year} {2015})}\BibitemShut
  {NoStop}%
\bibitem [{\citenamefont {Gonzalez}, \citenamefont {Cúñez},\ and\
  \citenamefont {Franklin}(2023)}]{Gonzalez}%
  \BibitemOpen
  \bibfield  {author} {\bibinfo {author} {\bibfnamefont {J.~O.}\ \bibnamefont
  {Gonzalez}}, \bibinfo {author} {\bibfnamefont {F.~D.}\ \bibnamefont
  {Cúñez}},\ and\ \bibinfo {author} {\bibfnamefont {E.~M.}\ \bibnamefont
  {Franklin}},\ }\bibfield  {title} {\enquote {\bibinfo {title} {Bidisperse
  beds sheared by viscous fluids: {G}rain segregation and bed hardening},}\
  }\href {https://doi.org/10.1063/5.0168415} {\bibfield  {journal} {\bibinfo
  {journal} {Phys. Fluids}\ }\textbf {\bibinfo {volume} {35}},\ \bibinfo
  {pages} {103326} (\bibinfo {year} {2023})}\BibitemShut {NoStop}%
\bibitem [{\citenamefont {Yalin}(1977)}]{Yalin_1}%
  \BibitemOpen
  \bibfield  {author} {\bibinfo {author} {\bibfnamefont {M.~S.}\ \bibnamefont
  {Yalin}},\ }\href@noop {} {\emph {\bibinfo {title} {Mechanics of sediment
  transport}}},\ \bibinfo {edition} {1st}\ ed.\ (\bibinfo  {publisher}
  {Pergamon Press},\ \bibinfo {year} {1977})\BibitemShut {NoStop}%
\bibitem [{\citenamefont {Allen}\ and\ \citenamefont
  {Kudrolli}(2018)}]{Allen_2}%
  \BibitemOpen
  \bibfield  {author} {\bibinfo {author} {\bibfnamefont {B.}~\bibnamefont
  {Allen}}\ and\ \bibinfo {author} {\bibfnamefont {A.}~\bibnamefont
  {Kudrolli}},\ }\bibfield  {title} {\enquote {\bibinfo {title} {Granular bed
  consolidation, creep, and armoring under subcritical fluid flow},}\ }\href
  {https://doi.org/10.1103/PhysRevFluids.3.074305} {\bibfield  {journal}
  {\bibinfo  {journal} {Phys. Rev. Fluids}\ }\textbf {\bibinfo {volume} {3}},\
  \bibinfo {pages} {074305} (\bibinfo {year} {2018})}\BibitemShut {NoStop}%
\bibitem [{\citenamefont {Jop}, \citenamefont {Forterre},\ and\ \citenamefont
  {Pouliquen}(2006)}]{Jop}%
  \BibitemOpen
  \bibfield  {author} {\bibinfo {author} {\bibfnamefont {P.}~\bibnamefont
  {Jop}}, \bibinfo {author} {\bibfnamefont {Y.}~\bibnamefont {Forterre}},\ and\
  \bibinfo {author} {\bibfnamefont {O.}~\bibnamefont {Pouliquen}},\ }\bibfield
  {title} {\enquote {\bibinfo {title} {A constitutive law for dense granular
  flows},}\ }\href@noop {} {\bibfield  {journal} {\bibinfo  {journal} {Nature}\
  }\textbf {\bibinfo {volume} {441}} (\bibinfo {year} {2006})}\BibitemShut
  {NoStop}%
\bibitem [{\citenamefont {Charru}, \citenamefont {Mouilleron-Arnould},\ and\
  \citenamefont {Eiff}(2004)}]{Charru_1}%
  \BibitemOpen
  \bibfield  {author} {\bibinfo {author} {\bibfnamefont {F.}~\bibnamefont
  {Charru}}, \bibinfo {author} {\bibfnamefont {H.}~\bibnamefont
  {Mouilleron-Arnould}},\ and\ \bibinfo {author} {\bibfnamefont
  {O.}~\bibnamefont {Eiff}},\ }\bibfield  {title} {\enquote {\bibinfo {title}
  {Erosion and deposition of particles on a bed sheared by a viscous flow},}\
  }\href@noop {} {\bibfield  {journal} {\bibinfo  {journal} {J. Fluid Mech.}\
  }\textbf {\bibinfo {volume} {519}},\ \bibinfo {pages} {55--80} (\bibinfo
  {year} {2004})}\BibitemShut {NoStop}%
\bibitem [{\citenamefont {Masteller}\ and\ \citenamefont
  {Finnegan}(2017)}]{Masteller_1}%
  \BibitemOpen
  \bibfield  {author} {\bibinfo {author} {\bibfnamefont {C.~C.}\ \bibnamefont
  {Masteller}}\ and\ \bibinfo {author} {\bibfnamefont {N.~J.}\ \bibnamefont
  {Finnegan}},\ }\bibfield  {title} {\enquote {\bibinfo {title} {Interplay
  between grain protrusion and sediment entrainment in an experimental
  flume},}\ }\href@noop {} {\bibfield  {journal} {\bibinfo  {journal} {J.
  Geophys. Res. Earth Surf.}\ }\textbf {\bibinfo {volume} {122}},\ \bibinfo
  {pages} {274--289} (\bibinfo {year} {2017})}\BibitemShut {NoStop}%
\bibitem [{\citenamefont {Masteller}\ \emph {et~al.}(2019)\citenamefont
  {Masteller}, \citenamefont {Finnegan}, \citenamefont {Turowski},
  \citenamefont {Yager},\ and\ \citenamefont {Rickenmann}}]{Masteller_2}%
  \BibitemOpen
  \bibfield  {author} {\bibinfo {author} {\bibfnamefont {C.~C.}\ \bibnamefont
  {Masteller}}, \bibinfo {author} {\bibfnamefont {N.~J.}\ \bibnamefont
  {Finnegan}}, \bibinfo {author} {\bibfnamefont {J.~M.}\ \bibnamefont
  {Turowski}}, \bibinfo {author} {\bibfnamefont {E.~M.}\ \bibnamefont
  {Yager}},\ and\ \bibinfo {author} {\bibfnamefont {D.}~\bibnamefont
  {Rickenmann}},\ }\bibfield  {title} {\enquote {\bibinfo {title}
  {History-dependent threshold for motion revealed by continuous bedload
  transport measurements in a steep mountain stream.}}\ }\href@noop {}
  {\bibfield  {journal} {\bibinfo  {journal} {Geophys. Res. Lett.}\ }\textbf
  {\bibinfo {volume} {46}},\ \bibinfo {pages} {2583--2591} (\bibinfo {year}
  {2019})}\BibitemShut {NoStop}%
\bibitem [{\citenamefont {Cates}\ \emph {et~al.}(1998)\citenamefont {Cates},
  \citenamefont {Wittmer}, \citenamefont {Bouchaud},\ and\ \citenamefont
  {Claudin}}]{Cates}%
  \BibitemOpen
  \bibfield  {author} {\bibinfo {author} {\bibfnamefont {M.~E.}\ \bibnamefont
  {Cates}}, \bibinfo {author} {\bibfnamefont {J.~P.}\ \bibnamefont {Wittmer}},
  \bibinfo {author} {\bibfnamefont {J.-P.}\ \bibnamefont {Bouchaud}},\ and\
  \bibinfo {author} {\bibfnamefont {P.}~\bibnamefont {Claudin}},\ }\bibfield
  {title} {\enquote {\bibinfo {title} {Jamming, force chains, and fragile
  matter},}\ }\href {https://doi.org/10.1103/PhysRevLett.81.1841} {\bibfield
  {journal} {\bibinfo  {journal} {Phys. Rev. Lett.}\ }\textbf {\bibinfo
  {volume} {81}},\ \bibinfo {pages} {1841--1844} (\bibinfo {year}
  {1998})}\BibitemShut {NoStop}%
\bibitem [{\citenamefont {Majmudar}\ and\ \citenamefont
  {Behringer}(2005)}]{Majmudar}%
  \BibitemOpen
  \bibfield  {author} {\bibinfo {author} {\bibfnamefont {T.~S.}\ \bibnamefont
  {Majmudar}}\ and\ \bibinfo {author} {\bibfnamefont {R.~P.}\ \bibnamefont
  {Behringer}},\ }\bibfield  {title} {\enquote {\bibinfo {title} {Contact force
  measurements and stress-induced anisotropy in granular materials},}\
  }\href@noop {} {\bibfield  {journal} {\bibinfo  {journal} {Nature}\ }\textbf
  {\bibinfo {volume} {435}},\ \bibinfo {pages} {1079--1082} (\bibinfo {year}
  {2005})}\BibitemShut {NoStop}%
\bibitem [{\citenamefont {Bi}\ \emph {et~al.}(2011)\citenamefont {Bi},
  \citenamefont {Zhang}, \citenamefont {Chakraborty},\ and\ \citenamefont
  {Behringer}}]{Bi}%
  \BibitemOpen
  \bibfield  {author} {\bibinfo {author} {\bibfnamefont {D.}~\bibnamefont
  {Bi}}, \bibinfo {author} {\bibfnamefont {J.}~\bibnamefont {Zhang}}, \bibinfo
  {author} {\bibfnamefont {B.}~\bibnamefont {Chakraborty}},\ and\ \bibinfo
  {author} {\bibfnamefont {R.~P.}\ \bibnamefont {Behringer}},\ }\bibfield
  {title} {\enquote {\bibinfo {title} {Jamming by shear},}\ }\href@noop {}
  {\bibfield  {journal} {\bibinfo  {journal} {Nature}\ }\textbf {\bibinfo
  {volume} {480}},\ \bibinfo {pages} {355--358} (\bibinfo {year}
  {2011})}\BibitemShut {NoStop}%
\bibitem [{\citenamefont {C\'u\~nez}\ \emph {et~al.}(2022)\citenamefont
  {C\'u\~nez}, \citenamefont {Franklin}, \citenamefont {Houssais},
  \citenamefont {Arratia},\ and\ \citenamefont {Jerolmack}}]{Cunez2}%
  \BibitemOpen
  \bibfield  {author} {\bibinfo {author} {\bibfnamefont {F.~D.}\ \bibnamefont
  {C\'u\~nez}}, \bibinfo {author} {\bibfnamefont {E.~M.}\ \bibnamefont
  {Franklin}}, \bibinfo {author} {\bibfnamefont {M.}~\bibnamefont {Houssais}},
  \bibinfo {author} {\bibfnamefont {P.}~\bibnamefont {Arratia}},\ and\ \bibinfo
  {author} {\bibfnamefont {D.~J.}\ \bibnamefont {Jerolmack}},\ }\bibfield
  {title} {\enquote {\bibinfo {title} {Strain hardening by sediment
  transport},}\ }\href@noop {} {\bibfield  {journal} {\bibinfo  {journal}
  {Phys. Rev. Res.}\ }\textbf {\bibinfo {volume} {4}},\ \bibinfo {pages}
  {L022055} (\bibinfo {year} {2022})}\BibitemShut {NoStop}%
\bibitem [{\citenamefont {Frey}\ and\ \citenamefont {Church}(2009)}]{Frey}%
  \BibitemOpen
  \bibfield  {author} {\bibinfo {author} {\bibfnamefont {P.}~\bibnamefont
  {Frey}}\ and\ \bibinfo {author} {\bibfnamefont {M.}~\bibnamefont {Church}},\
  }\bibfield  {title} {\enquote {\bibinfo {title} {How river beds move},}\
  }\href@noop {} {\bibfield  {journal} {\bibinfo  {journal} {Science}\ }\textbf
  {\bibinfo {volume} {325}},\ \bibinfo {pages} {1509--1510} (\bibinfo {year}
  {2009})}\BibitemShut {NoStop}%
\bibitem [{\citenamefont {Zhou}\ \emph {et~al.}(2020)\citenamefont {Zhou},
  \citenamefont {Cui}, \citenamefont {Jing}, \citenamefont {Zhao},
  \citenamefont {Song},\ and\ \citenamefont {Huang}}]{Zhou2}%
  \BibitemOpen
  \bibfield  {author} {\bibinfo {author} {\bibfnamefont {G.~G.~D.}\
  \bibnamefont {Zhou}}, \bibinfo {author} {\bibfnamefont {K.~F.~E.}\
  \bibnamefont {Cui}}, \bibinfo {author} {\bibfnamefont {L.}~\bibnamefont
  {Jing}}, \bibinfo {author} {\bibfnamefont {T.}~\bibnamefont {Zhao}}, \bibinfo
  {author} {\bibfnamefont {D.}~\bibnamefont {Song}},\ and\ \bibinfo {author}
  {\bibfnamefont {Y.}~\bibnamefont {Huang}},\ }\bibfield  {title} {\enquote
  {\bibinfo {title} {Particle size segregation in granular mass flows with
  different ambient fluids},}\ }\href
  {https://doi.org/https://doi.org/10.1029/2020JB019536} {\bibfield  {journal}
  {\bibinfo  {journal} {J. Geophys. Res.-Sol. Ea.}\ }\textbf {\bibinfo {volume}
  {125}},\ \bibinfo {pages} {e2020JB019536} (\bibinfo {year}
  {2020})}\BibitemShut {NoStop}%
\bibitem [{\citenamefont {Cui}, \citenamefont {Zhou},\ and\ \citenamefont
  {Jing}(2021)}]{Cui}%
  \BibitemOpen
  \bibfield  {author} {\bibinfo {author} {\bibfnamefont {K.~F.~E.}\
  \bibnamefont {Cui}}, \bibinfo {author} {\bibfnamefont {G.~G.~D.}\
  \bibnamefont {Zhou}},\ and\ \bibinfo {author} {\bibfnamefont
  {L.}~\bibnamefont {Jing}},\ }\bibfield  {title} {\enquote {\bibinfo {title}
  {Viscous effects on the particle size segregation in geophysical mass flows:
  Insights from immersed granular shear flow simulations},}\ }\href
  {https://doi.org/https://doi.org/10.1029/2021JB022274} {\bibfield  {journal}
  {\bibinfo  {journal} {J. Geophys. Res.-Sol. Ea.}\ }\textbf {\bibinfo {volume}
  {126}},\ \bibinfo {pages} {e2021JB022274} (\bibinfo {year}
  {2021})}\BibitemShut {NoStop}%
\bibitem [{\citenamefont {Rousseau}, \citenamefont {Chauchat},\ and\
  \citenamefont {Frey}(2022)}]{Rousseau}%
  \BibitemOpen
  \bibfield  {author} {\bibinfo {author} {\bibfnamefont {H.}~\bibnamefont
  {Rousseau}}, \bibinfo {author} {\bibfnamefont {J.}~\bibnamefont {Chauchat}},\
  and\ \bibinfo {author} {\bibfnamefont {P.}~\bibnamefont {Frey}},\ }\bibfield
  {title} {\enquote {\bibinfo {title} {Experiments on a single large particle
  segregating in bedload transport},}\ }\href
  {https://doi.org/10.1103/PhysRevFluids.7.064305} {\bibfield  {journal}
  {\bibinfo  {journal} {Phys. Rev. Fluids}\ }\textbf {\bibinfo {volume} {7}},\
  \bibinfo {pages} {064305} (\bibinfo {year} {2022})}\BibitemShut {NoStop}%
\bibitem [{\citenamefont {Frey}\ \emph {et~al.}(2020)\citenamefont {Frey},
  \citenamefont {{Lafaye de Micheaux}}, \citenamefont {Bel}, \citenamefont
  {Maurin}, \citenamefont {Rorsman}, \citenamefont {Martin},\ and\
  \citenamefont {Ducottet}}]{Frey2}%
  \BibitemOpen
  \bibfield  {author} {\bibinfo {author} {\bibfnamefont {P.}~\bibnamefont
  {Frey}}, \bibinfo {author} {\bibfnamefont {H.}~\bibnamefont {{Lafaye de
  Micheaux}}}, \bibinfo {author} {\bibfnamefont {C.}~\bibnamefont {Bel}},
  \bibinfo {author} {\bibfnamefont {R.}~\bibnamefont {Maurin}}, \bibinfo
  {author} {\bibfnamefont {K.}~\bibnamefont {Rorsman}}, \bibinfo {author}
  {\bibfnamefont {T.}~\bibnamefont {Martin}},\ and\ \bibinfo {author}
  {\bibfnamefont {C.}~\bibnamefont {Ducottet}},\ }\bibfield  {title} {\enquote
  {\bibinfo {title} {Experiments on grain size segregation in bedload transport
  on a steep slope},}\ }\href
  {https://doi.org/https://doi.org/10.1016/j.advwatres.2019.103478} {\bibfield
  {journal} {\bibinfo  {journal} {Adv. Water Resour.}\ }\textbf {\bibinfo
  {volume} {136}},\ \bibinfo {pages} {103478} (\bibinfo {year}
  {2020})}\BibitemShut {NoStop}%
\bibitem [{\citenamefont {Ferdowsi}\ \emph {et~al.}(2017)\citenamefont
  {Ferdowsi}, \citenamefont {Ortiz}, \citenamefont {Houssais},\ and\
  \citenamefont {Jerolmack}}]{Ferdowsi}%
  \BibitemOpen
  \bibfield  {author} {\bibinfo {author} {\bibfnamefont {B.}~\bibnamefont
  {Ferdowsi}}, \bibinfo {author} {\bibfnamefont {C.~P.}\ \bibnamefont {Ortiz}},
  \bibinfo {author} {\bibfnamefont {M.}~\bibnamefont {Houssais}},\ and\
  \bibinfo {author} {\bibfnamefont {D.~J.}\ \bibnamefont {Jerolmack}},\
  }\bibfield  {title} {\enquote {\bibinfo {title} {River-bed armouring as a
  granular segregation phenomenon},}\ }\href@noop {} {\bibfield  {journal}
  {\bibinfo  {journal} {Nat. Commun.}\ }\textbf {\bibinfo {volume} {8}}
  (\bibinfo {year} {2017})}\BibitemShut {NoStop}%
\bibitem [{\citenamefont {Kloss}\ and\ \citenamefont {Goniva}(2010)}]{Kloss}%
  \BibitemOpen
  \bibfield  {author} {\bibinfo {author} {\bibfnamefont {C.}~\bibnamefont
  {Kloss}}\ and\ \bibinfo {author} {\bibfnamefont {C.}~\bibnamefont {Goniva}},\
  }\bibfield  {title} {\enquote {\bibinfo {title} {{LIGGGHTS}: a new open
  source discrete element simulation software},}\ }in\ \href@noop {} {\emph
  {\bibinfo {booktitle} {Proc. 5th Int. Conf. on Discrete Element Methods}}}\
  (\bibinfo {address} {London, UK},\ \bibinfo {year} {2010})\BibitemShut
  {NoStop}%
\bibitem [{\citenamefont {Berger}\ \emph {et~al.}(2015)\citenamefont {Berger},
  \citenamefont {Kloss}, \citenamefont {Kohlmeyer},\ and\ \citenamefont
  {Pirker}}]{Berger}%
  \BibitemOpen
  \bibfield  {author} {\bibinfo {author} {\bibfnamefont {R.}~\bibnamefont
  {Berger}}, \bibinfo {author} {\bibfnamefont {C.}~\bibnamefont {Kloss}},
  \bibinfo {author} {\bibfnamefont {A.}~\bibnamefont {Kohlmeyer}},\ and\
  \bibinfo {author} {\bibfnamefont {S.}~\bibnamefont {Pirker}},\ }\bibfield
  {title} {\enquote {\bibinfo {title} {Hybrid parallelization of the {LIGGGHTS}
  open-source {DEM} code},}\ }\href@noop {} {\bibfield  {journal} {\bibinfo
  {journal} {Powder Technol.}\ }\textbf {\bibinfo {volume} {278}},\ \bibinfo
  {pages} {234--247} (\bibinfo {year} {2015})}\BibitemShut {NoStop}%
\bibitem [{\citenamefont {Goniva}\ \emph {et~al.}(2012)\citenamefont {Goniva},
  \citenamefont {Kloss}, \citenamefont {Deen}, \citenamefont {Kuipers},\ and\
  \citenamefont {Pirker}}]{Goniva}%
  \BibitemOpen
  \bibfield  {author} {\bibinfo {author} {\bibfnamefont {C.}~\bibnamefont
  {Goniva}}, \bibinfo {author} {\bibfnamefont {C.}~\bibnamefont {Kloss}},
  \bibinfo {author} {\bibfnamefont {N.~G.}\ \bibnamefont {Deen}}, \bibinfo
  {author} {\bibfnamefont {J.~A.~M.}\ \bibnamefont {Kuipers}},\ and\ \bibinfo
  {author} {\bibfnamefont {S.}~\bibnamefont {Pirker}},\ }\bibfield  {title}
  {\enquote {\bibinfo {title} {Influence of rolling friction on single spout
  fluidized bed simulation},}\ }\href@noop {} {\bibfield  {journal} {\bibinfo
  {journal} {Particuology}\ }\textbf {\bibinfo {volume} {10}},\ \bibinfo
  {pages} {582--591} (\bibinfo {year} {2012})}\BibitemShut {NoStop}%
\bibitem [{\citenamefont {Zhou}\ \emph {et~al.}(2010)\citenamefont {Zhou},
  \citenamefont {Kuang}, \citenamefont {Chu},\ and\ \citenamefont {Yu}}]{Zhou}%
  \BibitemOpen
  \bibfield  {author} {\bibinfo {author} {\bibfnamefont {Z.~Y.}\ \bibnamefont
  {Zhou}}, \bibinfo {author} {\bibfnamefont {S.~B.}\ \bibnamefont {Kuang}},
  \bibinfo {author} {\bibfnamefont {K.~W.}\ \bibnamefont {Chu}},\ and\ \bibinfo
  {author} {\bibfnamefont {A.~B.}\ \bibnamefont {Yu}},\ }\bibfield  {title}
  {\enquote {\bibinfo {title} {Discrete particle simulation of particle–fluid
  flow: model formulations and their applicability},}\ }\href@noop {}
  {\bibfield  {journal} {\bibinfo  {journal} {J. Fluid Mech.}\ }\textbf
  {\bibinfo {volume} {661}},\ \bibinfo {pages} {482–510} (\bibinfo {year}
  {2010})}\BibitemShut {NoStop}%
\bibitem [{\citenamefont {Tsuji}, \citenamefont {Tanaka},\ and\ \citenamefont
  {Ishida}(1992)}]{Tsuji}%
  \BibitemOpen
  \bibfield  {author} {\bibinfo {author} {\bibfnamefont {Y.}~\bibnamefont
  {Tsuji}}, \bibinfo {author} {\bibfnamefont {T.}~\bibnamefont {Tanaka}},\ and\
  \bibinfo {author} {\bibfnamefont {T.}~\bibnamefont {Ishida}},\ }\bibfield
  {title} {\enquote {\bibinfo {title} {Lagrangian numerical simulation of plug
  flow of cohesionless particles in a horizontal pipe},}\ }\href@noop {}
  {\bibfield  {journal} {\bibinfo  {journal} {Powder Technol.}\ }\textbf
  {\bibinfo {volume} {71}},\ \bibinfo {pages} {239--250} (\bibinfo {year}
  {1992})}\BibitemShut {NoStop}%
\bibitem [{\citenamefont {Tsuji}, \citenamefont {Kawaguchi},\ and\
  \citenamefont {Tanaka}(1993)}]{Tsuji2}%
  \BibitemOpen
  \bibfield  {author} {\bibinfo {author} {\bibfnamefont {Y.}~\bibnamefont
  {Tsuji}}, \bibinfo {author} {\bibfnamefont {T.}~\bibnamefont {Kawaguchi}},\
  and\ \bibinfo {author} {\bibfnamefont {T.}~\bibnamefont {Tanaka}},\
  }\bibfield  {title} {\enquote {\bibinfo {title} {Discrete particle simulation
  of two-dimensional fluidized bed},}\ }\href@noop {} {\bibfield  {journal}
  {\bibinfo  {journal} {Powder Technol.}\ }\textbf {\bibinfo {volume} {77}},\
  \bibinfo {pages} {79--87} (\bibinfo {year} {1993})}\BibitemShut {NoStop}%
\bibitem [{\citenamefont {Liu}\ \emph {et~al.}(2016)\citenamefont {Liu},
  \citenamefont {Liu}, \citenamefont {Fu},\ and\ \citenamefont {G.}}]{Liu}%
  \BibitemOpen
  \bibfield  {author} {\bibinfo {author} {\bibfnamefont {D.}~\bibnamefont
  {Liu}}, \bibinfo {author} {\bibfnamefont {X.}~\bibnamefont {Liu}}, \bibinfo
  {author} {\bibfnamefont {X.}~\bibnamefont {Fu}},\ and\ \bibinfo {author}
  {\bibfnamefont {W.}~\bibnamefont {G.}},\ }\bibfield  {title} {\enquote
  {\bibinfo {title} {Quantification of the bed load effects on turbulent
  open-channel flows},}\ }\href@noop {} {\bibfield  {journal} {\bibinfo
  {journal} {J. Geophys. Res. Earth Surf.}\ }\textbf {\bibinfo {volume}
  {121}},\ \bibinfo {pages} {767--789} (\bibinfo {year} {2016})}\BibitemShut
  {NoStop}%
\bibitem [{\citenamefont {Lima}\ \emph {et~al.}(2022)\citenamefont {Lima},
  \citenamefont {Assis}, \citenamefont {Alvarez},\ and\ \citenamefont
  {Franklin}}]{Lima2}%
  \BibitemOpen
  \bibfield  {author} {\bibinfo {author} {\bibfnamefont {N.~C.}\ \bibnamefont
  {Lima}}, \bibinfo {author} {\bibfnamefont {W.~R.}\ \bibnamefont {Assis}},
  \bibinfo {author} {\bibfnamefont {C.~A.}\ \bibnamefont {Alvarez}},\ and\
  \bibinfo {author} {\bibfnamefont {E.~M.}\ \bibnamefont {Franklin}},\
  }\bibfield  {title} {\enquote {\bibinfo {title} {Grain-scale computations of
  barchan dunes},}\ }\href {https://doi.org/10.1063/5.0121810} {\bibfield
  {journal} {\bibinfo  {journal} {Phys. Fluids}\ }\textbf {\bibinfo {volume}
  {34}},\ \bibinfo {pages} {123320} (\bibinfo {year} {2022})},\ \Eprint
  {https://arxiv.org/abs/https://doi.org/10.1063/5.0121810}
  {https://doi.org/10.1063/5.0121810} \BibitemShut {NoStop}%
\bibitem [{\citenamefont {Gondret}, \citenamefont {Lance},\ and\ \citenamefont
  {Petit}(2002)}]{Gondret}%
  \BibitemOpen
  \bibfield  {author} {\bibinfo {author} {\bibfnamefont {P.}~\bibnamefont
  {Gondret}}, \bibinfo {author} {\bibfnamefont {M.}~\bibnamefont {Lance}},\
  and\ \bibinfo {author} {\bibfnamefont {L.}~\bibnamefont {Petit}},\ }\bibfield
   {title} {\enquote {\bibinfo {title} {Bouncing motion of spherical particles
  in fluids},}\ }\href {https://doi.org/10.1063/1.1427920} {\bibfield
  {journal} {\bibinfo  {journal} {Phys. Fluids}\ }\textbf {\bibinfo {volume}
  {14}},\ \bibinfo {pages} {643--652} (\bibinfo {year} {2002})}\BibitemShut
  {NoStop}%
\bibitem [{\citenamefont {Roache}(1998)}]{roache1998verification}%
  \BibitemOpen
  \bibfield  {author} {\bibinfo {author} {\bibfnamefont {P.~J.}\ \bibnamefont
  {Roache}},\ }\href@noop {} {\emph {\bibinfo {title} {Verification and
  validation in computational science and engineering}}},\ Vol.\ \bibinfo
  {volume} {895}\ (\bibinfo  {publisher} {Hermosa Albuquerque, NM},\ \bibinfo
  {year} {1998})\BibitemShut {NoStop}%
\bibitem [{\citenamefont {Derakhshani}, \citenamefont {Schott},\ and\
  \citenamefont {Lodewijks}(2015)}]{Derakhshani}%
  \BibitemOpen
  \bibfield  {author} {\bibinfo {author} {\bibfnamefont {S.~M.}\ \bibnamefont
  {Derakhshani}}, \bibinfo {author} {\bibfnamefont {D.~L.}\ \bibnamefont
  {Schott}},\ and\ \bibinfo {author} {\bibfnamefont {G.}~\bibnamefont
  {Lodewijks}},\ }\bibfield  {title} {\enquote {\bibinfo {title} {Micro–macro
  properties of quartz sand: {E}xperimental investigation and {DEM}
  simulation},}\ }\href
  {https://doi.org/https://doi.org/10.1016/j.powtec.2014.08.072} {\bibfield
  {journal} {\bibinfo  {journal} {Powder Technol.}\ }\textbf {\bibinfo {volume}
  {269}},\ \bibinfo {pages} {127--138} (\bibinfo {year} {2015})}\BibitemShut
  {NoStop}%
\bibitem [{\citenamefont {Courant}, \citenamefont {Friedrichs},\ and\
  \citenamefont {Lewy}(1967)}]{Courant}%
  \BibitemOpen
  \bibfield  {author} {\bibinfo {author} {\bibfnamefont {R.}~\bibnamefont
  {Courant}}, \bibinfo {author} {\bibfnamefont {K.}~\bibnamefont
  {Friedrichs}},\ and\ \bibinfo {author} {\bibfnamefont {H.}~\bibnamefont
  {Lewy}},\ }\bibfield  {title} {\enquote {\bibinfo {title} {On the partial
  difference equations of mathematical physics},}\ }\href@noop {} {\bibfield
  {journal} {\bibinfo  {journal} {IBM J. Res. Dev.}\ }\textbf {\bibinfo
  {volume} {11}},\ \bibinfo {pages} {215--234} (\bibinfo {year}
  {1967})}\BibitemShut {NoStop}%
\bibitem [{\citenamefont {Gonzalez}\ and\ \citenamefont
  {Franklin}(2024{\natexlab{a}})}]{Supplemental}%
  \BibitemOpen
  \bibfield  {author} {\bibinfo {author} {\bibfnamefont {J.~O.}\ \bibnamefont
  {Gonzalez}}\ and\ \bibinfo {author} {\bibfnamefont {E.~M.}\ \bibnamefont
  {Franklin}},\ }\bibfield  {title} {\enquote {\bibinfo {title} {Forces and
  grain-grain contacts in bidisperse beds sheared by viscous fluids {I}:
  {P}re-{CFDEM} scripts},}\ }\href
  {https://doi.org/https://doi.org/10.17632/992j4h96ky} {\bibfield  {journal}
  {\bibinfo  {journal} {Mendeley Data}\ } (\bibinfo {year}
  {2024}{\natexlab{a}}),\ https://doi.org/10.17632/992j4h96ky}\BibitemShut
  {NoStop}%
\bibitem [{\citenamefont {Gonzalez}\ and\ \citenamefont
  {Franklin}(2024{\natexlab{b}})}]{Supplemental2}%
  \BibitemOpen
  \bibfield  {author} {\bibinfo {author} {\bibfnamefont {J.~O.}\ \bibnamefont
  {Gonzalez}}\ and\ \bibinfo {author} {\bibfnamefont {E.~M.}\ \bibnamefont
  {Franklin}},\ }\bibfield  {title} {\enquote {\bibinfo {title} {Forces and
  grain-grain contacts in bidisperse beds sheared by viscous fluids {II}:
  {D}ata\textunderscore 6rpm},}\ }\href
  {https://doi.org/https://doi.org/10.17632/nzt86s3r8p} {\bibfield  {journal}
  {\bibinfo  {journal} {Mendeley Data}\ } (\bibinfo {year}
  {2024}{\natexlab{b}}),\ https://doi.org/10.17632/nzt86s3r8p}\BibitemShut
  {NoStop}%
\bibitem [{\citenamefont {Houssais}\ \emph {et~al.}(2016)\citenamefont
  {Houssais}, \citenamefont {Ortiz}, \citenamefont {Durian},\ and\
  \citenamefont {Jerolmack}}]{Houssais_2}%
  \BibitemOpen
  \bibfield  {author} {\bibinfo {author} {\bibfnamefont {M.}~\bibnamefont
  {Houssais}}, \bibinfo {author} {\bibfnamefont {C.~P.}\ \bibnamefont {Ortiz}},
  \bibinfo {author} {\bibfnamefont {D.~J.}\ \bibnamefont {Durian}},\ and\
  \bibinfo {author} {\bibfnamefont {D.~J.}\ \bibnamefont {Jerolmack}},\
  }\bibfield  {title} {\enquote {\bibinfo {title} {Rheology of sediment
  transported by a laminar flow},}\ }\href@noop {} {\bibfield  {journal}
  {\bibinfo  {journal} {Phys. Rev. E}\ }\textbf {\bibinfo {volume} {94}},\
  \bibinfo {pages} {062609} (\bibinfo {year} {2016})}\BibitemShut {NoStop}%
\bibitem [{\citenamefont {GDR-MiDi}(2004)}]{GDR_midi}%
  \BibitemOpen
  \bibfield  {author} {\bibinfo {author} {\bibnamefont {GDR-MiDi}},\ }\bibfield
   {title} {\enquote {\bibinfo {title} {On dense granular flows},}\ }\href
  {https://doi.org/10.1140/epje/i2003-10153-0} {\bibfield  {journal} {\bibinfo
  {journal} {The European Physical Journal E}\ }\textbf {\bibinfo {volume}
  {14}} (\bibinfo {year} {2004}),\ 10.1140/epje/i2003-10153-0}\BibitemShut
  {NoStop}%
\bibitem [{\citenamefont {Jing}, \citenamefont {Kwok},\ and\ \citenamefont
  {Leung}(2017)}]{Jing}%
  \BibitemOpen
  \bibfield  {author} {\bibinfo {author} {\bibfnamefont {L.}~\bibnamefont
  {Jing}}, \bibinfo {author} {\bibfnamefont {C.~Y.}\ \bibnamefont {Kwok}},\
  and\ \bibinfo {author} {\bibfnamefont {Y.~F.}\ \bibnamefont {Leung}},\
  }\bibfield  {title} {\enquote {\bibinfo {title} {Micromechanical origin of
  particle size segregation},}\ }\href
  {https://doi.org/10.1103/PhysRevLett.118.118001} {\bibfield  {journal}
  {\bibinfo  {journal} {Phys. Rev. Lett.}\ }\textbf {\bibinfo {volume} {118}},\
  \bibinfo {pages} {118001} (\bibinfo {year} {2017})}\BibitemShut {NoStop}%
\bibitem [{\citenamefont {Radjai}\ \emph {et~al.}(1998)\citenamefont {Radjai},
  \citenamefont {Wolf}, \citenamefont {Jean},\ and\ \citenamefont
  {Moreau}}]{Radjai}%
  \BibitemOpen
  \bibfield  {author} {\bibinfo {author} {\bibfnamefont {F.}~\bibnamefont
  {Radjai}}, \bibinfo {author} {\bibfnamefont {D.~E.}\ \bibnamefont {Wolf}},
  \bibinfo {author} {\bibfnamefont {M.}~\bibnamefont {Jean}},\ and\ \bibinfo
  {author} {\bibfnamefont {J.-J.}\ \bibnamefont {Moreau}},\ }\bibfield  {title}
  {\enquote {\bibinfo {title} {Bimodal character of stress transmission in
  granular packings},}\ }\href {https://doi.org/10.1103/PhysRevLett.80.61}
  {\bibfield  {journal} {\bibinfo  {journal} {Phys. Rev. Lett.}\ }\textbf
  {\bibinfo {volume} {80}},\ \bibinfo {pages} {61--64} (\bibinfo {year}
  {1998})}\BibitemShut {NoStop}%
\end{thebibliography}%

\end{document}